\batchmode

\documentclass[10pt,english,reprint, aps, superscriptaddress, nofootinbib]{revtex4-2}
\usepackage[dvipsnames]{xcolor}
\usepackage[T1]{fontenc}
\usepackage[latin9]{inputenc}
\usepackage{babel}
\usepackage{verbatim}
\usepackage{amsbsy,bm}
\usepackage{amssymb}
\usepackage{esint}
\usepackage[normalem]{ulem}
\usepackage[pdftex,pdfusetitle,
 bookmarks=true,bookmarksnumbered=false,bookmarksopen=false,
 breaklinks=false,pdfborder={0 0 0},pdfborderstyle={},backref=false,colorlinks=true]
 {hyperref}
\hypersetup{
 linkcolor=blue, citecolor=cyan}

\makeatletter

\usepackage{xcolor}

\makeatother

\begin{document}
\title{Radiative corrections on vortical spin polarization in hot QCD matter}
\author{Shuo Fang}
\email{fangshuo@mail.ustc.edu.cn}

\affiliation{Department of Modern Physics, University of Science and Technology
of China, Anhui 230026, China}
\affiliation{Technical University of Munich, TUM School of Natural Sciences, Physics Department, James-Franck-Str. 1, 85748 Garching, Germany}

\author{Shi Pu}
\email{shipu@ustc.edu.cn}

\affiliation{Department of Modern Physics and Anhui Center for fundamental Sciences (Theoretical Physics), University of Science and Technology
of China, Anhui 230026, China}
\affiliation{Southern Center for Nuclear-Science Theory (SCNT), Institute of Modern
Physics, Chinese Academy of Sciences, Huizhou 516000, Guangdong Province,
China}
\author{Di-Lun Yang}
\email{dilunyang@gmail.com}

\affiliation{Institute of Physics, Academia Sinica, Taipei, 11529, Taiwan}
\affiliation{Physics Division, National Center for Theoretical Sciences, Taipei, 106319, Taiwan}
\begin{abstract}
We investigate the radiative corrections on spin polarization of relativistic fermions induced by vortical fields in thermal-equilibrium QCD matter at weak coupling. Such corrections stem from the self-energy gradients in quantum kinetic theory, which are further obtained by a more systematic and general approach through the Keldysh equation. By applying the hard-thermal-loop approximation, we obtain new corrections upon the spin-polarization spectrum and also the axial-charge current in connection to the axial/chiral vortical effect for massive quarks up to the leading order of the QCD coupling. Further influence on spin alignment of vector mesons from similar effects is also analyzed.   
\end{abstract}
\maketitle

\section{Introduction}
Relativistic heavy-ion collisions produce hot, dense matter governed by strong interactions, known as the quark-gluon plasma (QGP) \citep{Gyulassy:2004vg,Shuryak:2004cy}. This setting provides a platform for studying the properties of quantum chromodynamics (QCD) matter under extreme conditions, such as strong electromagnetic fields \citep{STAR:2023jdd,Huang:2015oca, Gao:2020vbh}, strong vorticity \citep{STAR:2017ckg,Becattini:2020ngo,Becattini:2024uha}, and anomalous transport \citep{Kharzeev:2015znc}. One of the promising probes in this context is the spin observables. The spin polarization of hyperons reveals that the QGP is a fast-rotating system \citep{STAR:2017ckg,Karpenko:2016jyx}, while its momentum spectrum indicates an interplay between the fine structure of local vorticity and shear stress within the QGP \citep{STAR:2019erd,Becattini:2021iol,Fu:2021pok,Yi:2021ryh, Wu:2022mkr}. Recent measurements of vector-meson spin alignment \citep{ALICE:2019aid,STAR:2022fan} further underscore the influence of strong force fields and their correlations \citep{Sheng:2022wsy,Sheng:2023urn, Muller:2021hpe, Kumar:2023ghs}.

Based on quantum field theory, quantum kinetic theory (QKT) \citep{Vasak:1987um, Son:2012zy, Chen:2012ca, Gao:2012ix, Hidaka:2016yjf, Gao:2019znl, Weickgenannt:2019dks, Weickgenannt:2020aaf, Yang:2020hri,Hidaka:2022dmn}
in terms of Wigner functions provides a unified framework to depict
the microscopic dynamics of various degrees of freedom of particles,
including particle number and spin. The perturbative solution of QKT can directly describe spin effects in the form of modified Cooper-Frye formulae \citep{Becattini:2013fla,Fang:2016vpj} and the spin density matrix \citep{Becattini:2020sww,Wagner:2022gza}, provided the spin kinetic equations are solved. For example,
in local equilibrium the spin polarization induced by vorticity and shear viscous tensor
are explicitly derived \citep{Hidaka:2017auj,Yi:2021ryh}.  

Despite the success in probing macroscopic properties such as the vorticity of the QGP through global polarization, an intriguing and important question remains: How can we probe the microscopic properties associated with the strong interactions of QCD matter through spin transport? Related studies have investigated the spin alignment of vector mesons influenced by fluctuating glasma fields \cite{Kumar:2022ylt,Kumar:2023ghs,Yang:2024qpy}, turbulent color fields \cite{Muller:2021hpe,Yang:2021fea}, and vector-meson fields \cite{Sheng:2019kmk,Sheng:2022wsy,Sheng:2023urn}. Moreover, the non-equilibrium corrections dependent on interactions for local spin-polarization spectra have been extensively explored in recent years \cite{Hidaka:2017auj,Yang:2018lew,Weickgenannt:2022zxs,Weickgenannt:2022qvh,Wagner:2024fry,Fang:2022ttm,Fang:2023bbw,Fang:2024vds,Lin:2022tma,Lin:2024zik,Lin:2024svh,Wang:2024lis}. However, even under equilibrium conditions, interaction-dependent corrections on spin polarization may exist, as suggested by the radiative corrections to the axial-charge current under vorticity for massless fermions \cite{Golkar:2012kb,Hou:2012xg}, known as the axial/chiral vortical effect (AVE/CVE) \cite{Vilenkin:1979ui,Son:2009tf,Landsteiner:2011cp}. In QKT, such radiative corrections could potentially stem from self-energy corrections rather than collisions at weak coupling \cite{Yamamoto:2023okm,Fang:2023bbw}. Nonetheless, obtaining a complete perturbative solution for general spin-$j$ Wigner functions is technically challenging \cite{Hattori:2020gqh}, especially concerning their off-equilibrium and interaction corrections. 

In this paper, we investigate radiative corrections to the spin polarization spectrum induced by vorticity in thermal equilibrium, utilizing chiral kinetic theory for massless fermions and a novel approach based on the Keldysh equation for fermions of arbitrary mass. We further implement this theoretical framework to study quarks in a weakly coupled QCD environment using the hard-thermal-loop (HTL) approximation and also explore the axial-charge current. A similar study employing different methods and approximations for QED plasmas has been presented in Ref.~\cite{Buzzegoli:2021jeh}.

The paper is organized as follows: 
We first review the chiral kinetic
theory with self-energy corrections and propose the global equilibrium
form of the interaction-corrected chiral Wigner functions in Sec.~\ref{sec:Chiral-kinetic-theory}.
Then we propose the Keldysh equation framework as an off-equilibrium
analogue of Schwinger-Dyson equation in the phase-space in Sec.~\ref{sec:Keldysh-equations-for}.
After that we present the follow-up applications of the Keldysh equation,
calculating radiative corrections on spin polarization and AVE for massive fermions in Sec.~\ref{sec:Fermion-spin-polarization}
and similar contribution to vector-meson spin alignment in Sec.~\ref{sec:Vector-meson-spin}.
A summary and outlook are given in Sec.~\ref{sec:Summary}.

We utilize the natural unit by setting $\hbar=c=k_{\mathrm{B}}=1$.
We work in the Minkowski spacetime with mostly minus metric $\eta^{\mu\nu}=\mathrm{diag}(1,-1,-1,-1)$
and use the notation $\epsilon^{0123}=\epsilon^{123}=+1$. We introduce
the (anti-)symmetric symbol 
\begin{eqnarray}
    A_{(\mu\nu)}=\frac{A_{\mu\nu}+A_{\nu\mu}}{2}, \;\;
    A_{[\mu\nu]}=\frac{A_{\mu\nu}-A_{\nu\mu}}{2}, \nonumber
\end{eqnarray}
and the projection
operator $\Delta^{\mu\nu}=\eta^{\mu\nu}-u^{\mu}u^{\nu}$ with $u^\mu$ a normalized timelike vector.

\section{Chiral kinetic theory and its global equilibrium form \label{sec:Chiral-kinetic-theory}}

The gauge-invariant Wigner functions for chiral fermions, with $\chi=\pm$
representing right- (R) and left-handed (L), respectively, are defined as \citep{Vasak:1987um, Chen:2012ca, Gao:2012ix, Hidaka:2016yjf,Hidaka:2022dmn,Yamamoto:2023okm}
\begin{eqnarray}
\mathcal{W}_{\chi}^{<}(q,X) & = & \int{\rm d}^{4}Ye^{-iq\cdot Y}U(x,y)\langle\psi_{\chi}^{\dagger}(y)\psi_{\chi}(x)\rangle,\label{eq:Def_CWF}
\end{eqnarray}
where $x=X+\frac{Y}{2}$ and $y=X-\frac{Y}{2}$ and $U(x,y)$ is a
straight line gauge-link \citep{Vasak:1987um}. The bracket denotes
the ensemble average for a given initial-time density operator \citep{Blaizot:2001nr}.
The Wigner functions and self-energies are decomposed into the basis
of Pauli matrices\citep{Hidaka:2016yjf}, e.g. $\mathcal{W}_{{\rm R}}^{<}=\overline{\sigma}_{\mu}\mathcal{W}_{{\rm R}}^{<,\mu}$.
In this work we mainly focus on the collisionless limit so that the
lessor/greater self-energies are expected to be zero, $\Sigma^{\lessgtr}=0$.
By solving the equations of motion (EoMs) under quasi-particle approximations,
we are able to obtain the perturbative solution of Eq.~(\ref{eq:Def_CWF})
up to $\mathcal{O}(\hbar^{1},\overline{\Sigma}_{\chi,\mu}^{1})$ \citep{Yamamoto:2023okm},
\begin{eqnarray}
 \mathcal{W}_{\chi}^{<,\mu}(q,X)& = & 2\pi\epsilon(q_{0})\left[\delta(\widetilde{q}^{2})\left(\widetilde{q}^{\mu}+\chi\hbar\widetilde{S}_{(n)}^{\mu\nu}\widetilde{\Delta}_{\nu}\right)\right.\nonumber \\
 &  + &\left. \frac{\chi\hbar}{2}\delta^{\prime}(\widetilde{q}^{2})\epsilon^{\mu\nu\rho\sigma}\widetilde{q}_{\nu}(F_{\rho\sigma}+2\Delta_{[\rho}\overline{\Sigma}_{\chi,\sigma]})\right]f_{\chi}, \nonumber\\
 \label{eq:Collisionless_CWF}
\end{eqnarray}
where $f_{\chi}$ is the chiral distribution function and $\epsilon(x)\equiv x/|x|$.
Here we have denoted 
\begin{eqnarray}
    \widetilde{q}^{\mu}&=&q^{\mu}-\overline{\Sigma}_{\chi}^{\mu}, \nonumber \\
    \overline{\Sigma}_{\chi}^{\mu}&=&\Sigma_{\chi}^{\delta,\mu}+{\rm Re}\Sigma_{\chi}^{{\rm r},\mu},
\end{eqnarray}
where $\Sigma_{\chi}^{\delta,\mu}$ is one-point potential and $\Sigma_{\chi}^{{\rm r},\mu}$ is the retarded self-energy.
We define the frame-dependent modified spin tensor 
\begin{eqnarray}
\widetilde{S}_{(n)}^{\mu\nu}=\epsilon^{\mu\nu\rho\sigma}\widetilde{q}_{\rho}n_{\sigma}/(2\widetilde{q}\cdot n),   
\end{eqnarray}
where $n^{\mu}$ is a space-time dependent frame vector depending on the choice of a spin basis. For
convenience, we also introduce the modified covariant derivative
\begin{eqnarray}
    \widetilde{\Delta}_{\mu}&=& \Delta_{\mu}+(\Delta_{\lambda}\overline{\Sigma}_{\chi,\mu})\partial_{q}^{\lambda}-(\partial_{q,\lambda}\overline{\Sigma}_{\chi,\mu})\partial^{\lambda}, \nonumber \\
    \Delta_{\mu}&=&\partial_{\mu}-F_{\mu\nu}\partial_{q}^{\nu},
\end{eqnarray}
where the derivative $\partial_\mu=\partial_{X,\mu}$ operates only on the space coordinate $X$ and
the $F_{\rho\sigma}$ is the field strength of background electromagnetic fields (EMF).

The accompanied chiral kinetic equation (CKE) \citep{Yamamoto:2023okm} delineating the dynamical evolution of $f_{\chi}$ up to $\mathcal{O}(\hbar^{1},\overline{\Sigma}_{\chi,\mu}^{1})$ reads  
\begin{eqnarray}
0 & = & \left\{ \widetilde{q}^{\alpha}+\chi\hbar\frac{S_{\widetilde{q}}^{\mu\alpha}}{\widetilde{q}\cdot n}\left[E_{\mu}+2n^{\beta}(\Delta_{[\mu}\overline{\Sigma}_{\chi,\beta]})-\widetilde{q}^{\beta}(\widetilde{\Delta}_{\mu}n_{\beta})\right]\right.\nonumber \\
&  & \left.+\chi\hbar\frac{\epsilon^{\mu\alpha\rho\sigma}\widetilde{q}_{\rho}(\widetilde{\Delta}_{\mu}n_{\sigma})}{2\widetilde{q}\cdot n}+\chi\hbar S_{\widetilde{q}}^{\mu\alpha}\widetilde{\Delta}_{\mu}\right\}\widetilde{\Delta}_{\alpha}f_{\chi},\label{eq:EoM_CDist}
\end{eqnarray}
where the on-shell condition $\widetilde{q}^{2}=-\chi\hbar\widetilde{S}_{(n)}^{\mu\nu}\widetilde{\Delta}_{\mu}\widetilde{q}_{\nu}$
is imposed implicitly and $E_{\mu}=F_{\mu\nu}n^{\nu}$ as an electric field defined in terms of the frame vector $n^{\mu}$. Although the frame vectors are involved in both the Wigner function in Eq.~(\ref{eq:Collisionless_CWF}) and the kinetic equation in Eq.~(\ref{eq:EoM_CDist}), they are in fact frame independent. Note that $f_{\chi}$ with quantum corrections also follows a modified frame transformation. Consequently, the physical results obtained from the QKT are frame independent.

Our goal is to derive the lowest-order interaction corrections to the Wigner functions, beyond the EMF effects. These corrections, known to originate from the medium background, are calculated by $\overline{\Sigma}_{\chi} \sim \mathcal{O}(g^{2})$, with collision contributions starting at $\mathcal{O}(g^{4})$ \citep{Fang:2023bbw}. However, deriving these corrections is generally challenging, as the self-energies modify not only the momentum and on-shell conditions but also the chiral distribution functions. In this paper, we make an early-stage effort to derive such corrections at least in global equilibrium. These corrections are crucial for further understanding the local spin polarization of $\Lambda$ hyperons \citep{STAR:2019erd,STAR:2023eck} and provide an alternative method to derive interaction corrections to anomalous transport effects, such as the chiral separation effects induced by vorticity 
\citep{Son:2009tf,Landsteiner:2011cp,Gao:2012ix}.

In global equilibrium, at the leading order in $\hbar$ and the first order
in $\overline{\Sigma}_{\chi}$, the chiral distribution function satisfies
\begin{eqnarray}
\widetilde{q}\cdot\widetilde{\Delta}f_{\chi}^{(0)} & = & 0,\label{eq:LO_CKE}
\end{eqnarray}
where the on-shell condition is not imposed. Since $f_{\chi}^{(0)}$
should satisfy the detailed balance of classical collision terms and
give conservation of energy-momentum and particle number, it should
only be a function of $\beta u\cdot q-\beta\mu_{\chi}$ with two Lagrangian
multipliers $\{\beta^\mu=\beta{u^\mu},\mu_{\chi}\}$. Here $\beta=T^{-1}$ and $\mu_{\chi}$ correspond to the inverse temperature and a chiral chemical potential of thermodynamics, respectively \citep{Landau:1980mil}. In this paper, we work in the fluid rest frame such that $u^{\mu}=(1,{\bm u})$ with $\bm u\approx 0$.
In this case, we should have
\begin{eqnarray}
f_{\chi}^{(0)} & = & \frac{1}{e^{\beta(u\cdot q-\mu_{\chi})}+1}.\label{eq:LO_CDist}
\end{eqnarray}
Then the Vlasov-type kinetic equation (\ref{eq:LO_CKE}) reduces to
\begin{eqnarray}
0 & = & \Bigl\{ q^{\mu}\Bigl[\beta (u\cdot\partial\overline{\Sigma}_{\chi,\mu})-\overline{\Sigma}_{\chi,\nu}(\partial^{\nu}\beta u_{\mu})-(\partial_{q,\nu}\overline{\Sigma}_{\chi,\mu})(\partial^{\nu}\beta u\cdot q)\nonumber \\
 &  & \;+(\partial_{q,\nu}\overline{\Sigma}_{\chi,\mu})\left((\partial^{\nu}\beta\mu_{\chi})+\beta E^{\nu}\right)-\left(\beta E_{\mu}+(\partial_{\mu}\beta\mu_{\chi})\right)\Bigr]\nonumber \\
 &  & +q^{\mu}q^{\nu}(\partial_{\mu}\beta u_{\nu})+\overline{\Sigma}_{\chi,\nu}\left((\partial^{\nu}\beta\mu_{\chi})+\beta E^{\nu}\right)\Bigr\}\partial_{\beta u\cdot q}f_{\chi}^{(0)},\nonumber \\
\end{eqnarray} 
from which one can read out the following equilibrium conditions by extracting the vanishing coefficients for independent bases, $\{1,q^{\mu},q^{\mu}q^{\nu}\}$:
\begin{eqnarray}
\partial^{\nu}(\beta\mu_{\chi})+\beta E^{\nu}=0,\nonumber \\ 
\partial_{(\mu}(\beta u_{\nu)})=0, \;\label{eq:Geq_cond_1}
\end{eqnarray}
and 
\begin{eqnarray}
\beta u\cdot\partial\overline{\Sigma}_{\chi,\mu}-\overline{\Sigma}_{\chi,\nu}\partial^{\nu}(\beta u_{\mu})-(\partial_{q,\nu}\overline{\Sigma}_{\chi,\mu})\partial^{\nu}(\beta u\cdot q)=0.\nonumber\\ \label{eq:Geq_cond_2}
\end{eqnarray}
Equation~(\ref{eq:Geq_cond_1}) is the well-known global equilibrium conditions
\citep{Gao:2012ix}, while Eq.~(\ref{eq:Geq_cond_2}) gives a constraint
for the self-energies in global equilibrium. Notably, the $\overline{\Sigma}_{\chi,\mu}$ obtained from the HTL approximation satisfies the constraint.

To obtain the $\mathcal{O}(\hbar^{1})$ equilibrium distribution function,
we demand the frame independence of Wigner functions \citep{Chen:2015gta,Hidaka:2016yjf}.
The distribution function can further be decomposed to a frame-independent
part $\delta f_{\chi}^{(u)}$ and a frame-dependent part $\delta f_{\chi}^{(n)}$,
i.e. $f_{\chi}^{(1)}=\delta f_{\chi}^{(u)}+\delta f_{\chi}^{(n)}$,
where the former can be an arbitrary correction, while the latter is
proposed to be
\begin{eqnarray}
\delta f_{\chi}^{(n)} & = & \frac{\hbar\chi}{\widetilde{q}\cdot u}\widetilde{S}_{(n)}^{\rho\nu}u_{[\nu}\widetilde{\Delta}_{\rho]}f_{\chi}^{(0)}.
\end{eqnarray}
Using Schouten's identity and Eq.~(\ref{eq:LO_CKE}), we find
\begin{eqnarray}
\delta(\widetilde{q}^{2})\widetilde{q}^{\mu}\delta f_{\chi}^{(n)} & = & -\delta(\widetilde{q}^{2})\hbar\chi\left(\widetilde{S}_{(n)}^{\mu\rho}-\widetilde{S}_{(u)}^{\mu\rho}\right)\widetilde{\Delta}_{\rho}f_{\chi}^{(0)}.
\end{eqnarray}
Therefore, the $\mathcal{O}(\hbar^{1})$ sector of chiral Wigner functions
now becomes frame-independent, 
\begin{eqnarray}
 &  & \delta(\widetilde{q}^{2})\left(\widetilde{q}^{\mu}f_{\chi}^{(1)}+\chi\hbar\widetilde{S}_{(n)}^{\mu\nu}\widetilde{\Delta}_{\nu}f_{\chi}^{(0)}\right)\nonumber \\
 & & = \delta(\widetilde{q}^{2})\left(\widetilde{q}^{\mu}\delta f_{\chi}^{(u)}+\chi\hbar\widetilde{S}_{(u)}^{\mu\nu}\widetilde{\Delta}_{\nu}f_{\chi}^{(0)}\right).
\end{eqnarray}
Albeit no general guideline to determine $\delta f_{\chi}^{(u)}$,
we may propose a generalization of the non-interacting case \citep{Chen:2015gta,Hidaka:2017auj},
\begin{eqnarray}
\delta f_{\chi}^{(u)} & = & \frac{\chi\hbar}{2}(\partial_{\beta q\cdot u}f_{\chi}^{(0)})\widetilde{S}_{(u)}^{\rho\nu}\Omega_{\rho\nu},\label{eq:NLO_CDist}
\end{eqnarray}
where we have defined the thermal vorticity 
\begin{eqnarray} \Omega^{\alpha\beta}=\partial^{[\alpha}(\beta u^{\beta]}).
\end{eqnarray}
Then in the absence of external EMF, the equilibrium chiral Wigner
function (\ref{eq:Collisionless_CWF}) now becomes 
\begin{eqnarray}
 & &\mathcal{W}_{\chi}^{<,\mu}(q,X)\nonumber \\ 
 &  =& 2\pi\epsilon(q_{0})\Bigl\{\delta(q^{2})\Bigl[(q^{\mu}-\overline{\Sigma}_{\chi}^{\mu})\nonumber \\
 &   & -\frac{\chi\hbar}{4}\epsilon^{\mu\alpha\rho\nu}(q_{\alpha}-\overline{\Sigma}_{\chi,\alpha})\Omega_{\rho\nu}\partial_{\beta q\cdot u}\Bigr]\nonumber \\
 &  & -\delta^{\prime}(q^{2})\Bigl[q\cdot\overline{\Sigma}_{\chi}\left(2q^{\mu}-\frac{\chi\hbar}{2}\epsilon^{\mu\rho\nu\alpha}q_{\alpha}\Omega_{\rho\nu}\partial_{\beta q\cdot u}\right)\nonumber \\
 &  & \;-\chi\hbar\epsilon^{\mu\nu\rho\sigma}q_{\nu}(\partial_{\rho}\overline{\Sigma}_{\chi,\sigma})\Bigr]\Bigr\} f_{\chi}^{(0)},\label{eq:Equi_CWF}
\end{eqnarray}
where we have kept the terms up to $\mathcal{O}(\overline{\Sigma}_{\chi,\mu}^{1})$
and used the equilibrium conditions (\ref{eq:Geq_cond_1}, \ref{eq:Geq_cond_2}).
To verify whether Eq.~(\ref{eq:NLO_CDist}) indeed represents the equilibrium distribution function, we insert the equilibrium chiral distribution function,
\begin{eqnarray}
f_{\chi}  =  \frac{1}{e^{\beta g}+1}, \
g =  u \cdot q - \mu_{\chi} + \frac{\chi \hbar}{2} \widetilde{S}_{(u)}^{\rho\nu} \Omega_{\rho\nu}, \label{eq:Full_CDist}
\end{eqnarray}
into the CKE in Eq.~(\ref{eq:EoM_CDist}) with $n^{\mu}=u^{\mu}$, we find that the equation is satisfied up to $\mathcal{O}(\hbar^{1}, \overline{\Sigma}_{\chi,\mu}^{1})$.
Furthermore, we will derive such an equilibrium Wigner function (\ref{eq:Equi_CWF}) using the Keldysh equation formalism in the subsequent sections.

\section{Keldysh equations for Wigner functions \label{sec:Keldysh-equations-for}}

Deriving the explicit form of the self-energy correction on the Wigner function from QKT is nontrivial, particularly when extending to massive fermions and vector bosons (including photons and vector mesons). More critically, generalizing these results to the near-equilibrium (local-equilibrium) scenario becomes even more complex, even if collision effects are disregarded. Furthermore, in strongly coupled systems, the quasi-particle paradigm is no longer applicable, rendering perturbative solutions of the Wigner functions unattainable.
In this section, we propose an alternative approach parallel to QKT: the Keldysh equation \citep{maciejko2007introduction} in phase space, which represents an integral-equation parallel of the kinetic equation manifested as a differential-integral equation. The Keldysh-equation formalism enables systematic incorporation of interaction and off-equilibrium corrections and it can be extended beyond the quasi-particle limit.

We start from the time-ordered fermionic two-point Green's function
along the Kadanoff-Baym (KB) contour
\begin{eqnarray}
 &  & S_{ab}(x,y)=\langle T_{\mathrm{c}}\psi_{a}(x)\overline{\psi}_{b}(y)\rangle\nonumber \\
 & = & \theta_{\mathrm{c}}(x_{0}-y_{0})S_{ab}^{>}(x,y)-\theta_{\mathrm{c}}(y_{0}-x_{0})S_{ab}^{<}(x,y),\label{eq:Def_Fermion_WF}
\end{eqnarray}
where $T_{{\rm c}},\theta_{{\rm c}}$ are the time-ordering operator
and Heaviside step function in the KB contour. We further denote $\langle...\rangle_{0}$
as the ensemble average under a general reference density operator
$\hat{\rho}_{0}$, which is taken as the non-interacting density operator in this work, and the corresponding correlation functions
as $\mathcal{O}_{0}$. For
example, in the global equilibrium case it takes the form of \citep{Becattini:2012tc,Becattini:2014yxa}
\begin{eqnarray}
\hat{\rho}_{0} & = & \frac{\exp\left[-\beta_{\mu}\hat{P}_{0}^{\mu}+\alpha\hat{Q}_{0}+\varpi_{\mu\nu}\hat{\mathcal{J}}_{0}^{\mu\nu}\right]}{{\rm Tr}\exp\left[-\beta_{\mu}\hat{P}_{0}^{\mu}+\alpha\hat{Q}_{0}+\varpi_{\mu\nu}\hat{\mathcal{J}}_{0}^{\mu\nu}\right]},\label{eq:Geq_DO}
\end{eqnarray}
where $\hat{P}^{\mu},\hat{Q},\hat{\mathcal{J}}^{\mu\nu}$ are the
momentum operator, U(1) charge operator, and total angular momentum
operator of the system, respectively. Accordingly, $\hat{H}_{0}=\hat{P}^{0}_{0}$ and $\{\beta_{\mu},\alpha,\varpi_{\mu\nu}\}$
are the Lagrangian multipliers of the conserved charges of the system;
the subscript ``0'' denotes the free operators. We also assume that
the interaction information in the full density operator $\hat{\rho}$
only lies in the interaction sector of Hamiltonian $\hat{H}_{{\rm int}}=\hat{P}^{0}_{{\rm int}}$,
e.g. this corresponds to set $\hat{\mathcal{J}}_{{\rm int}}^{\mu\nu}=0$ in the global equilibrium density operator (\ref{eq:Geq_DO}), which only
holds for $\varpi_{\mu\nu}=\varpi_{ij}\eta_{\mu}^{i}\eta_{\nu}^{j}$
\citep{Weinberg:1995mt}. 
We thus work in the real-time formalism and obtain the following
Keldysh equation \citep{maciejko2007introduction}, 
\begin{eqnarray}
 &  & S^{<}(x,y)=S_{0}^{<}(x,y)-\int_{-\infty}^{+\infty}\mathrm{d}^{4}z_{1}\mathrm{d}^{4}z_{2}\left[S^{\mathrm{r}}(x,z_{2})\Sigma^{\mathrm{r}}(z_{2},z_{1})\right.\nonumber \\
 &  & \;\times S_{0}^{<}(z_{1},y)+S^{\mathrm{r}}(x,z_{2})\Sigma^{<}(z_{2},z_{1})S_{0}^{\mathrm{a}}(z_{1},y)\nonumber \\
 &  & \;\left.+S^{<}(x,z_{2})\Sigma^{\mathrm{a}}(z_{2},z_{1})S_{0}^{\mathrm{a}}(z_{1},y)\right]-i\int_{-\infty}^{+\infty}\mathrm{d}^{4}z_{1}\nonumber \\
 &  & \;\left(S^{\mathrm{r}}(x,z_{1})\Sigma(z_{1})S_{0}^{<}(z_{1},y)+S^{<}(x,z_{1})\Sigma(z_{1})S_{0}^{\mathrm{a}}(z_{1},y)\right),\label{eq:Fermion_Keldysh_Eq_1}
\end{eqnarray}
where the two point self-energy and one-point potential are defined
as
\begin{eqnarray}
\Sigma_{cd}(z_{2},z_{1}) & = & \left\langle T_{\mathrm{c}}i^{2}\int_{\mathrm{c}}\mathrm{d}^{4}z^{\prime}\mathrm{d}^{4}z\frac{\delta\mathcal{L}_{\mathrm{int}}(z^{\prime})}{\delta\overline{\psi}_{c}(z_{2})}\frac{\delta\mathcal{L}_{\mathrm{int}}(z)}{\delta\psi_{d}(z_{1})}\right\rangle,\;\\
\Sigma_{cd}(z_{1}) & = & \left\langle T_{\mathrm{c}}i\int_{\mathrm{c}}\mathrm{d}^{4}z\frac{\delta\mathcal{L}_{\mathrm{int}}(z)}{\delta\overline{\psi}_c(z_{1})\delta\psi_d(z_{1})}\right\rangle.
\end{eqnarray}
Here the subscripts of the fermion fields represent the spinor indices.
Thus $\Sigma_{cd}(z_{1},z_{2})$ does not acquire a singular part as in Ref.~\citep{Blaizot:2001nr}
and can also be written as Eq.~(\ref{eq:Def_Fermion_WF}), 
\begin{eqnarray}
 &  & \Sigma_{cd}(z_{1},z_{2})\nonumber \\
 & = & \theta_{\mathrm{c}}(t_{1}-t_{2})\Sigma_{cd}^{>}(z_{1},z_{2})-\theta_{\mathrm{c}}(t_{2}-t_{1})\Sigma_{cd}^{<}(z_{1},z_{2}).
\end{eqnarray}
Here we have introduced the (anti-)Feynman Green's functions and the
retarded/advanced (r/a) quantities for the Fermi-type correlators,
\begin{eqnarray}
 &  & S^{\mathrm{f}}(x,y)=\left[S^{\mathrm{\overline{f}}}(x,y)\right]^{\dagger}\nonumber \\
 & = & \theta(x_{0}-y_{0})S^{>}(x,y)-\theta(y_{0}-x_{0})S^{<}(x,y),
\end{eqnarray}
and
\begin{eqnarray}
\mathcal{O}_{\mathrm{r}}(x,y) & = & i\theta(x_{0}-y_{0})\left(\mathcal{O}^{>}(x,y)+\mathcal{O}^{<}(x,y)\right),\\
\mathcal{O}_{\mathrm{a}}(x,y) & = & -i\theta(y_{0}-x_{0})\left(\mathcal{O}^{>}(x,y)+\mathcal{O}^{<}(x,y)\right),
\end{eqnarray}
where $\mathcal{O}$ can be self-energies or Green's functions. For the bose-type correlators, we can simply change the minus sign in front
of $\mathcal{O}^{<}$ to the plus sign as opposed to the Grassmann nature
of fermionic fields. It is easy to obtain the Keldysh equations in
terms of Wigner functions, 
\begin{eqnarray}
 &  & S^{<}(q,X)\nonumber \\
 & = & S_{0}^{<}(q,X)-\left(S^{\mathrm{r}}(q,X)\star\Sigma^{\mathrm{r}}(q,X)\right)\star S_{0}^{<}(q,X)\nonumber \\
 &  & -\left(S^{\mathrm{r}}(q,X)\star\Sigma^{<}(q,X)\right)\star S_{0}^{\mathrm{a}}(q,X)\nonumber \\
 &  & -\left(S^{<}(q,X)\star\Sigma^{\mathrm{a}}(q,X)\right)\star S_{0}^{\mathrm{a}}(q,X)\nonumber \\
 &  & -i\left(S^{\mathrm{r}}(q,X)\star\Sigma(X)\right)\star S_{0}^{<}(q,X)\nonumber \\
 &  & -i\left(S^{<}(q,X)\star\Sigma(X)\right)\star S_{0}^{\mathrm{a}}(q,X),\label{eq:FWF_Keldysh_eq_1}
\end{eqnarray}
where 
\begin{eqnarray}
H(q,X)  =  \left[f(q,X)\star g(q,X)\right]\star h(q,X)\label{eq:Three-Moyal-product}
\end{eqnarray}
stems from the Wigner transform of
\begin{eqnarray}
H(x,y) & = & \int{\rm d}^{4}z_{1}{\rm d}^{4}z_{2}f(x,z_{1})g(z_{1},z_{2})h(z_{2},y)\label{eq:Three-Moyal-product-coordinate-space}
\end{eqnarray}
and the Moyal product with associative properties is defined as 
\begin{eqnarray}
f\star g & \equiv & f\exp\left(-i\hbar\frac{\overleftarrow{\partial}\cdot\overrightarrow{\partial}_{q}-\overrightarrow{\partial}\cdot\overleftarrow{\partial}_{q}}{2}\right)g.\label{eq:Moyal_product}
\end{eqnarray}
Equations~(\ref{eq:Three-Moyal-product}) and (\ref{eq:Three-Moyal-product-coordinate-space}) also hold for the situation involved with one-point potential, where $g(z_1,z_2)=g(z_1)\delta^{(4)}(z_1-z_2)$ and $g(q,X)=g(X)$.

Similarly, for the vector fields, we have 
\begin{eqnarray}
 &  & G_{\mu\nu}^{<}(q,X)\nonumber \\
 & = & G_{\mu\nu}^{<,0}(q,X)-(G_{\mu\beta}^{\mathrm{r}}(q,X)\star\Sigma^{\beta\alpha,\mathrm{r}}(q,X))\star G_{\alpha\nu}^{0,<}(q,X)\nonumber \\
 &  & \;-(G_{\mu\beta}^{\mathrm{r}}(q,X)\star\Sigma^{\beta\alpha,<}(q,X))\star G_{\alpha\nu}^{0,\mathrm{a}}(q,X)\nonumber \\
 &  & \;-(G_{\mu\beta}^{<}(q,X)\star\Sigma^{\beta\alpha,\mathrm{a}}(q,X))\star G_{\alpha\nu}^{0,\mathrm{a}}(q,X),\label{eq:BWF_Keldysh_eq}
\end{eqnarray}
which originates from the Wigner transform of the lessor two point function of a general complex
vector field
\begin{eqnarray}
G_{\mu\nu}^{<}(x,y) & = & \langle A_{\nu}^{\dagger}(y)A_{\mu}(x)\rangle,
\end{eqnarray}
and the greater function is $G_{\nu\mu}^{>}(y,x)=G_{\nu\mu}^{<,\dagger}(x,y)$.
These Wigner functions are symmetric w.r.t. the Lorentz indices for
the real parts and anti-symmetric for imaginary parts. Namely,
the anti-symmetric part of $G_{\mu\nu}^{<}(p,X)$ is purely imaginary,
which can be verified by evaluating the photonic Wigner functions
\citep{Hattori:2020gqh,Fang1:2025}.

We emphasize again that the Wigner functions with ``0'' subscripts
in Eqs.~(\ref{eq:FWF_Keldysh_eq_1}, \ref{eq:BWF_Keldysh_eq}) can be general out-of-equilibrium Wigner functions without interaction corrections. 
More concrete examples can be found in Refs.~\citep{Li:2025pef, Fang2:2025}.
These Wigner functions serve as the initial input in the Keldysh-equation
formalism and can be derived from statistical field theory \citep{DeGroot:1980dk,Becattini:2013fla,Palermo:2021hlf,Becattini:2021suc},
linear response theory \citep{Liu:2020dxg,Liu:2021uhn}, and the collisionless
QKT \citep{Gao:2012ix,Hidaka:2022dmn}.
Additionally, their validity can be further verified by quantum kinetic equations (QKEs) with vanishing vector and axial collision kernels \citep{Weickgenannt:2020aaf,Wang:2020pej,Fang:2022ttm}. 
Notably, some of us have developed a method to calculate
the global equilibrium Wigner functions up to an arbitrarily high order in gradient or $\hbar$ expansions
easily from the generalized KMS condition \citep{Fang1:2025}.
In this work, we use the global equilibrium Wigner functions with
a general profile of fluid velocity $u^{\mu}$ for simplicity.

As an integral-equation parallel to QKT, the Wigner functions derived from
Eqs.~(\ref{eq:FWF_Keldysh_eq_1}, \ref{eq:BWF_Keldysh_eq}) should
be equivalent to solving QKEs, provided $S_{0}^{<},G_{\mu\nu,0}^{<}$
are appropriately specified.
A key advantage of this formalism is the relative simplicity with which interaction corrections can be included. The use of derivatives in the Moyal product combined with the spacetime dependence of the Wigner functions and self-energies facilitates the execution of the gradient expansion and perturbation in the coupling constant order-by-order. At the leading order in gradients, we also expect that the results can be reproduced by linear response theory. Our goal
in this work is to derive the lowest-order interaction corrections
to fermionic spin-polarization pseudovector and spin density matrix elements
within such a formalism.

\section{Fermionic spin polarization and axial vortical effects \label{sec:Fermion-spin-polarization} }

In this section, we derive the interaction corrections to the fermionic
spin-polarization pseudo-vector up to the leading order in gradients and
lowest order in the coupling constant. 
We will first present the reproduction of the well-known polarization vector in a background U(1) field in Sec.~\ref{subsec:Contribution-from-U(1)}. Subsequently, we will explore the implications in a QCD background in Sec.~\ref{subsec:Contribution-from-QCD}, ensuring consistency with the results from CKT presented in Sec.~\ref{sec:Chiral-kinetic-theory} and detailed in Ref.~\citep{Fang:2023bbw}.

\subsection{Contribution from U(1) background field  \label{subsec:Contribution-from-U(1)} }

In the presence of a background U(1) field, up to the leading order
in the coupling constant, we may drop the two-point self-energies and
the Keldysh equation for fermions (\ref{eq:FWF_Keldysh_eq_1}) reduces
to
\begin{eqnarray}
 &  & S^{<}(q,X)\nonumber \\
 & = & S_{0}^{<}(q)-gA_{\mu}(X)\left[S_{0}^{\mathrm{r}}(q)\gamma^{\mu}S_{0}^{<}(q)+S_{0}^{<}(q)\gamma^{\mu}S_{0}^{\mathrm{a}}(q)\right]\nonumber \\
 &  & +i\hbar\partial_{\nu}gA_{\mu}(X)\Bigl[S_{0}^{\mathrm{r}}(q)\frac{\overrightarrow{\partial}_{q}^{\nu}-\overleftarrow{\partial}_{q}^{\nu}}{2}\gamma^{\mu}S_{0}^{<}(q)\nonumber \\
 &  & \;+S_{0}^{<}(q)\frac{\overrightarrow{\partial}_{q}^{\nu}-\overleftarrow{\partial}_{q}^{\nu}}{2}\gamma^{\mu}S_{0}^{\mathrm{a}}(q)\Bigr]+\mathcal{O}(g^{2},\partial^{2}),\label{eq:FWF_Keq_U1bkg_1}
\end{eqnarray}
where we have taken $\Sigma(X)=-igA_{\mu}(X)\gamma^{\mu}$ with $A_{\mu}(X)$ being
a spacetime-dependent Abelian background field and we have used the spacetime-independent
 fermionic r/a Wigner functions in equilibrium (see also Appendix.\ref{subsec:Non-interacting-equilibrium-Wign} for more details).
 If we omit the hydrodynamic gradient for simplicity, the global
equilibrium fermionic Wigner function $S_{0}^{<}(q,X)$  acquires a
simple form \citep{Bellac:2011kqa}, 
\begin{eqnarray}
S_{0}^{<}(q) & = & 2\pi\epsilon(q_{0})\delta(q^{2}-m^{2})f_{\mathrm{f}}^{<}(q_{0})(\gamma^{\alpha}q_{\alpha}+m),\label{eq:Geq_Fermi_WF_O(1)}
\end{eqnarray}
where the fermionic distribution function is Fermi-Dirac type $f_{\mathrm{f}}^{<}(q_{0})=1/[\exp(\beta (q_{0}-\mu))+1]$ with $\mu$ being a U(1) vector-charge chemical potential. Inserting Eqs.~(\ref{eq:Geq_Fermi_WF_O(1)}) and (\ref{eq:r/a_FWF_global_eq}) into Eq.~(\ref{eq:FWF_Keq_U1bkg_1}),
we obtain
\begin{eqnarray}\label{eq:S_up_to_g}
    S^{<}(q,X) & = & S_{0}^{<}(q)-2\pi\epsilon(q_{0})\delta^{\prime}(q^{2}-m^{2})f_{\mathrm{f}}^{<}(q_{0})\nonumber\\
 &  & \;\times\{gA_{\mu}(X)(\gamma^{\alpha}\gamma^{\mu}\gamma^{\beta}q_{\beta}q_{\alpha}+2mq^{\mu}+m^{2}\gamma^{\mu})\nonumber\\
 &  & \;+\hbar\frac{ig}{2}\partial_{\nu}A_{\mu}(X)[(\gamma^{\nu}\gamma^{\mu}\gamma^{\alpha}-\gamma^{\alpha}\gamma^{\mu}\gamma^{\nu})q_{\alpha}\nonumber\\
 &  & \;+m(\gamma^{\nu}\gamma^{\mu}-\gamma^{\mu}\gamma^{\nu})]\},
\end{eqnarray}
where we have focused on the global equilibrium case with  $\partial_{i}A_{j}\neq0$ only. In the above calculations, we have used the following
properties of the regularized Dirac Delta function \citep{Landsman:1986uw,Bellac:2011kqa}
\begin{eqnarray}
\frac{1}{x+i\epsilon}\delta(x) & = & -\frac{1}{2}\delta^{\prime}(x)-i\pi\left(\delta(x)\right)^{2},\label{eq:Regularized_delta_property_1}
\end{eqnarray}
thus the superficial divergence terms related to $\left[\delta(q^{2}-m^{2})\right]^{2}$
are canceled automatically. Using the gamma matrices decomposition
\begin{eqnarray*}
\gamma^{\mu}\gamma^{\nu}\gamma^{\rho} & = & -i\epsilon^{\mu\nu\rho\sigma}\gamma^{5}\gamma_{\sigma}+\eta^{\mu\nu}\gamma^{\rho}+\eta^{\nu\rho}\gamma^{\mu}-\eta^{\mu\rho}\gamma^{\nu},
\end{eqnarray*}
we derive the $S^<(q,X)$ up to $\mathcal{O}(g^{1},\partial^{1})$
\begin{eqnarray}
     &  & S^{<}(q,X)\nonumber\\
 & = & 2\pi\epsilon(q_{0})f_{\mathrm{f}}^{<}(q_{0})\nonumber\\
 &  & \;\times\{\delta[(q-gA)^{2}-m^{2}][\gamma^{\alpha}(q_{\alpha}-gA_{\alpha})+m]\nonumber\\
 &  & \;+\delta^{\prime}(q^{2}-m^{2})\frac{\hbar}{2}gF_{\mu\nu}(\epsilon^{\mu\nu\rho\alpha}\gamma^{5}\gamma_{\rho}q_{\alpha}-m\gamma^{\mu\nu})\},\label{eq:FWF_U1BkgF}
\end{eqnarray}
where the field strength tensor is defined as $F_{\alpha\beta}=2\partial_{[\alpha}A_{\beta]}$
and $\gamma^{\mu\nu}=i\gamma^{[\mu}\gamma^{\nu]}$. Eq.~(\ref{eq:FWF_U1BkgF})
is consistent with those derived from QKEs \citep{Fang:2016vpj,Hattori:2019ahi}.
We notice that the canonical momentum $q_{\mu}$ is modified to the
kinetic one $q_{\mu}-gA_{\mu}$ in Eq. (\ref{eq:FWF_U1BkgF}), which is  also observed in the previous studies  \citep{Mrowczynski:1992hq,Yamamoto:2023okm,Fang:2023bbw}. 
This alteration ensures the gauge invariance of Wigner functions. More precisely, the $q_{\mu}-gA_{\mu}$ here should correspond to $q^{\mu}$ in Eq.~(\ref{eq:Collisionless_CWF}).

The background-field-induced axial-vector and tensor components
of the fermionic Wigner functions, proportional to $\gamma^5\gamma_{\rho}$ and $\gamma^{\mu\nu}$, respectively,  manifest that the fermions are polarized by
an external field. We also emphaize that the background field $A_{\mu}$ can be EMF or other effective meson fields. 
The nontrivial correlation from background effective meson fields can explain the global spin alignment of $\phi$ mesons measured by STAR collaboration \citep{Sheng:2019kmk,Sheng:2022wsy,STAR:2022fan}.

\subsection{Contribution from QCD background  \label{subsec:Contribution-from-QCD} and the corrections to spin polarization}

We now consider the Wigner functions for massive quarks under a thermal QCD background. 
In our analysis, the one-point function for chromo-electromagnetic fields is set to zero, whereas the gluonic correlation functions remain non-vanishing.  Consequently, the interaction-dependent corrections that we consider in this section are limited to terms of order $\mathcal{O}(g^2)$.
We work in the Clifford
components of Wigner functions by decomposing
\begin{eqnarray}
S^{\lessgtr} & = & \mathcal{F}^{\lessgtr}+i\mathcal{P}^{\lessgtr}\gamma^{5}+\gamma^{\mu}\mathcal{V}_{\mu}^{\lessgtr}+\gamma^{5}\gamma^{\mu}\mathcal{A}_{\mu}^{\lessgtr}+\frac{1}{2}\gamma^{\mu\nu}\mathcal{S}_{\mu\nu}^{\lessgtr},\nonumber \\
\end{eqnarray}
where $\mathcal{F}^{\lessgtr}$, $\mathcal{P}^{\lessgtr}$, $\mathcal{V}_{\mu}^{\lessgtr}$, $\mathcal{A}_{\mu}^{\lessgtr}$, and $\mathcal{S}_{\mu\nu}^{\lessgtr}$ correspond to the scalar, pseudo-scalar, vector, axial-vector, and tensor components of Wigner functions, respectively. For simplicity, we focus on the color-averaged quantities and thus neglect the color decomposition.
These components have well-established power counting by assuming
spin effects are relatively small, i.e. $\mathcal{F},\mathcal{V}^{\mu}\sim\mathcal{O}(\partial^{0})$,
$\mathcal{A}^{\mu},\mathcal{S}^{\mu\nu}\sim\mathcal{O}(\partial^{1})$
and $\mathcal{P}\sim\mathcal{O}(\partial^{2})$ (also see e.g. Refs.~\citep{Hattori:2019ahi,Yang:2020hri,Fang:2023bbw}).

The self-energies can also be decomposed with the same manner. 
The imaginary part of r/a self-energies contributes to the collision
kernel, while the real parts plays a similar role as background fields
\citep{Hidaka:2016yjf,Yamamoto:2023okm,Fang:2023bbw}. The collision
effects are at least of $\mathcal{O}(g^{4})$ \citep{Arnold:2000dr}
and $\Sigma^{\lessgtr}$ at one-loop level corresponds to the splitting
or collinear 2$\to$3 processes in gauge theories \citep{Arnold:2002zm,Arnold:2003zc}. Therefore, in the lowest order in coupling constant, we only need to focus on the real
parts of r/a self-energy in one-loop, i.e.
\begin{eqnarray}
\rm Re\,\Sigma^{\mathrm{r}} & = & \rm Re\,\Sigma^{\mathrm{a}}=-\mathrm{Im}\,\Sigma_{\mathrm{f}},\label{eq:LO_rSE}
\end{eqnarray}
where $\mathrm{Im}\,\Sigma$ denotes imaginary parts of the Clifford
components. Up to $\mathcal{O}(g^{2})$, 
\begin{eqnarray}
\Sigma_{\mathrm{f}}(q,X) & = & -C_{\mathrm{F}}g^{2}\int\frac{\mathrm{d}^{4}p}{(2\pi)^{4}}G_{\mathrm{f}}^{\mu\nu}(q-p,X)\gamma_{\mu}S_{\mathrm{f}}(p,X)\gamma_{\nu},\nonumber \\
\end{eqnarray}
where we have factorized the color factor $t_{a}t_{a}=C_{\mathrm{F}}$
so that the gluonic propagator reduces to the photonic propagator. We
insert the photonic and fermionic Wigner functions  up to the first
order in gradients, whose expression can be found in Appendix.\ref{subsec:Non-interacting-equilibrium-Wign} and in Refs. 
~\citep{Hattori:2019ahi,Hattori:2020gqh}, into the above equation for $\Sigma_{\mathrm{f}}(q,X)$. We find that in the HTL approximation and up to $\mathcal{O}(\partial^{1})$, only the vector and axial-vector sectors of $\Sigma_{\mathrm{f}}(q,X)$ are non-zero,
\begin{eqnarray}
\mathrm{Im}\,\Sigma_{\mathrm{f}}^{\mathrm{V},\beta} & = & -m_{\mathrm{f}}^{2}\left[\frac{u^{\beta}}{|\boldsymbol{q}|}Q_{0}\left(\frac{q_{0}}{|\boldsymbol{q}|}\right)+\frac{q_{\perp}^{\beta}}{|\boldsymbol{q}|^{2}}Q_{1}\left(\frac{q_{0}}{|\boldsymbol{q}|}\right)\right],\label{eq:Vector_SE}\\
\mathrm{Im}\,\Sigma_{\mathrm{f}}^{\mathrm{A},\beta} & = & -\frac{C_{\mathrm{F}}g^{2}T\mu}{16\pi^{2}m_{\rm f}^2}\hbar\epsilon^{\beta\nu\rho\sigma}\Omega_{\rho\sigma}\mathrm{Im}\Sigma_{\mathrm{f},\nu}^{\mathrm{V}}.\label{eq:Axial_SE}
\end{eqnarray}
Here we choose $u^{\mu}$ as the fluid velocity of the QCD background and introduce
fermionic thermal mass $m_{\mathrm{f}}^{2}=\frac{C_{\mathrm{F}}g^{2}}{8}\left(T^{2}+\frac{\mu^{2}}{\pi^{2}}\right)$.
We decompose $q^\mu$ as 
\begin{eqnarray}
    q^{\mu}&=&q_{0}u^{\mu}+q_{\perp}^{\mu}, \nonumber \\
    q_{0}&=&u\cdot q,\;\;q_{\perp}^{\mu}=\Delta^{\mu\nu}q_{\nu},
\end{eqnarray}
and define 
\begin{eqnarray}
    |\boldsymbol{q}|=\sqrt{-q_{\perp}\cdot q_{\perp}},\;
    \hat{q}_{\perp}^{\mu}=q_{\perp}^{\mu}|\boldsymbol{q}|^{-1}.
\end{eqnarray}
The auxiliary functions $Q(x)$ are defined as
\begin{eqnarray}
Q_{0}(x)=\frac{1}{2}\ln\left|\frac{x+1}{x-1}\right| & ,\; & Q_{1}(x)=xQ_{0}(x)-1.
\end{eqnarray}

Taking the axial-vector Clifford component, $\mathcal{A}^{<,\mu}$, in Eq.~(\ref{eq:FWF_Keldysh_eq_1})
and expanding it up to the first order in gradient, we obtain
\begin{eqnarray}
 &  & \mathcal{A}^{<,\mu}(q,X)\nonumber \\
 & = & \mathcal{A}_{0}^{<,\mu}(q,X)+2\pi\epsilon(q_{0})\delta(q^{2}-m^{2})f_{\mathrm{f}}^{<}f_{\mathrm{f}}^{>}\nonumber \\
 &  & \;\times\frac{\hbar}{4}\epsilon^{\mu\rho\alpha\beta}\Omega_{\alpha\beta}{\rm Im}\Sigma_{{\rm f},\rho}^{{\rm V}}+2\pi\epsilon(q_{0})\delta^{\prime}(q^{2}-m^{2})\hbar\nonumber \\
 &  & \;\Bigl\{\left[\Omega_{\alpha\beta}q\cdot{\rm Im}\Sigma_{{\rm f}}^{{\rm V}}+{\rm Im}\Sigma_{{\rm f},\beta}^{{\rm V}}(q^{\lambda}\xi_{\alpha\lambda}-(\partial_{\alpha}\alpha))\right]\nonumber \\
 &  & \;\times\frac{1}{2}\epsilon^{\mu\rho\alpha\beta}q_{\rho}f_{\mathrm{f}}^{<}f_{\mathrm{f}}^{>}+\Bigl[-\epsilon^{\mu\rho\alpha\beta}q_{\rho}(\partial_{\alpha}{\rm Im}\Sigma_{{\rm f},\beta}^{{\rm V}})\nonumber \\
 &  & \;-2q^{\mu}q\cdot{\rm Im}\Sigma_{{\rm f}}^{{\rm A}}+(m^{2}+q^{2}){\rm Im}\Sigma_{{\rm f}}^{{\rm A},\mu}\Bigr]f_{\mathrm{f}}^{<}\Bigr\},\label{eq:AV_WF_HTLa}
\end{eqnarray}
where  $f_{{\rm f}}^{>}=1-f_{{\rm f}}^{<}$ and we have introduced the 
thermal chemical potential $\alpha=\beta \mu$ and thermal shear tensor $\xi_{\alpha\beta}$ 
\begin{eqnarray}
   \xi_{\alpha\beta}=\partial_{(\alpha}(\beta u_{\beta)}).
\end{eqnarray}
To derive the expression of $\mathcal{A}^{<,\mu}$, we have substituted equilibrium Wigner functions and carried out
the following calculation with Eq.~(\ref{eq:Regularized_delta_property_1}),
\begin{eqnarray}
\left[\mathcal{F}_{0}^{{\rm r}}+\mathcal{F}_{0}^{{\rm a}}\right]\delta(q^{2}-m^{2}) & = & m\delta^{\prime}(q^{2}-m^{2}),
\end{eqnarray}
where $\mathcal{F}_{0}^{{\rm r}}$ and $\mathcal{F}_{0}^{{\rm a}}$ are listed below Eq.~(\ref{eq:r/a_FWF_global_eq}).
Such a derivative applied to the Dirac delta function signifies corrections to the on-shell conditions induced by interactions, similar to those in Eq.~(\ref{eq:S_up_to_g}). Notably, in the off-shell part, the term $\epsilon^{\mu\rho\alpha\beta}q_{\rho}\partial_{\alpha}{\rm Im}\Sigma_{{\rm f},\beta}^{{\rm V}}$ agrees with the findings in Ref.~\citep{Fang:2023bbw}. 
The off-global-equilibrium
contribution $\partial_\mu \alpha$ and $\xi_{\alpha\beta}$ in Eq.~(\ref{eq:AV_WF_HTLa}) originates from the anisotropy of the fluid field. 

Let us discuss the application to the spin polarization. We will focus on the global equilibrium case for fermionic spin polarization and AVE for simplicity. Integrating over $q_{0}$ of $\mathcal{A}^{<, \mu}$ in Eq.~(\ref{eq:AV_WF_HTLa}) and taking the normalization, we can obtain
a full QCD interaction correction up to $\mathcal{O}(g^{2})$ to the spin modified Cooper-Frye formulae \citep{Becattini:2013fla,Fang:2016vpj}. 
These new corrections include the remaining terms from the so-called dynamical contributions neglected in our previous work \citep{Fang:2023bbw}. Here, for simplicity, we list the new corrections to the spin polarization pseudo-vector,
\begin{eqnarray}
\delta\mathcal{P}_{\mathrm{therm}}^{\mu}(t,\boldsymbol{q}) & = & -\frac{1}{4mN}\int_{\Sigma}\frac{m_{{\rm f}}^{2}}{E_{\boldsymbol{q}}^{2}}G_{{\rm T}}(\boldsymbol{q})\epsilon^{\mu\nu\alpha\beta}q_{\nu}\Omega_{\alpha\beta},\label{eq:C-F_formulae_therm}\\
\delta\mathcal{P}_{\mathrm{acc}}^{\mu}(t,\boldsymbol{q}) & = & \frac{1}{8mN}\int_{\Sigma}\frac{m_{{\rm f}}^{2}}{E_{\boldsymbol{q}}^{3}}G_{{\rm a}}(\boldsymbol{q})\epsilon^{\mu\nu\rho\sigma}q_{\rho}u_{\sigma}(u\cdot\partial)u_{\nu},\nonumber \\
\label{eq:C-F_formulae_acc}\\\delta\mathcal{P}_{\mathrm{vor}}^{\mu}(t,\boldsymbol{q}) & = & \frac{1}{4mN}\int_{\Sigma}\frac{m_{{\rm f}}^{2}}{E_{\boldsymbol{q}}^{2}}\Bigg[\omega^{\mu}G_{{\rm vor}1}(\boldsymbol{q})\nonumber \\
 &  & \;-\frac{\omega\cdot q}{E_{\boldsymbol{q}}}\left(u^{\mu}G_{{\rm vor}2}(\boldsymbol{q})+\frac{q_{\perp}^{\mu}}{E_{q}}G_{{\rm vor}3}(\boldsymbol{q})\right)\Bigg],\nonumber\\
 \label{eq:C-F_formulae_vor}
\end{eqnarray}
where we define 
\begin{eqnarray}
    \int_{\Sigma} = \int_\Sigma q\cdot d\sigma,
\end{eqnarray}
with $\Sigma$ being the chemical freeze-out hypersurface and  ${\rm d}\sigma$
being the volume elements. The $N$ is given by,
\begin{eqnarray}
    N=\int_{\Sigma} f_{\mathrm{f}}^{<}(E_{\boldsymbol{q}},X).
\end{eqnarray} 
The particles are set to be on-shell, thus $E_{\boldsymbol{q}}=\sqrt{\boldsymbol{q}^{2}+m^{2}}$. The lower index "therm", "acc" and "vor" means the corrections 
to the spin polarization induced by the thermal vorticity, fluid acceleration 
and kinematic vorticity, respectively.
The auxiliary functions, such as $G_{\rm T}({\bm q})$, $G_{{\rm a}}({\bm q})$, etc., for contributions from thermal vorticity, acceleration, and kinematic vorticity are listed in Appendix \ref{sec:Auxiliary-functions-in}. 

Now, we comment on Eqs.~(\ref{eq:C-F_formulae_therm}, \ref{eq:C-F_formulae_acc},
\ref{eq:C-F_formulae_vor}). We emphasize that we have worked in the $\mu=0$ and strictly global equilibrium case,  characterized by a finite vorticity tensor 
\begin{eqnarray}    \omega_{\alpha\beta}=\Delta_{\mu\alpha}\Delta_{\nu\beta}\partial^{[\mu}{u^{\nu]}},
\end{eqnarray}and a finite fluid vorticity vector
\begin{eqnarray}
    \omega^\mu =\frac{1}{2} \epsilon^{\mu\nu\rho\sigma}u_\nu \omega_{\rho\sigma}.
\end{eqnarray}
Therefore, these contributions exist even for global polarization. Different with our previous work \citep{Yi:2021ryh, Wu:2022mkr, Fang:2023bbw, Fang:2024vds} in the massless case, the corrections to the spin polarization acquired above is for the massive quark. A  smooth connection to the massless limit is straightforward to get. For phenomenological application, the so-called strange-equilibrium scenario may be adopted with the assumption that the polarization spectrum of the strange quark is smoothly connected to the one for the corresponding $\Lambda$ hyperon \cite{Liang:2004ph, Fu:2021pok,Yi:2021ryh}.

As a cross check, we will recover the interaction modified chiral
Wigner function (\ref{eq:Equi_CWF}) and equilibrium distribution
(\ref{eq:Full_CDist}) in the strict global equilibrium case, by imposing
the Killing equation (\ref{eq:Geq_cond_1}) which gives $\xi_{\mu\nu}=\partial_{\mu}\alpha=0$.
Working in the massless limit and assuming zero chemical potential,
Eq.~(\ref{eq:AV_WF_HTLa}) reduces to
\begin{eqnarray}
 &  & \mathcal{A}^{<,\mu}(q,X)\nonumber \\
 & = & \mathcal{A}_{0}^{<,\mu}-2\pi\epsilon(q_{0})\delta(q^{2})f_{\mathrm{f}}^{<}f_{\mathrm{f}}^{>}\frac{\hbar}{4}\epsilon^{\mu\rho\alpha\beta}\Omega_{\alpha\beta}\overline{\Sigma}_{{\rm V},\rho}\nonumber \\
 &  & \;+2\pi\epsilon(q_{0})\delta^{\prime}(q^{2})\hbar\Bigl\{\epsilon^{\mu\rho\alpha\beta}q_{\rho}(\partial_{\alpha}\overline{\Sigma}_{{\rm V},\beta})f_{\mathrm{f}}^{<}\nonumber \\
 &  & \;-\Omega_{\alpha\beta}q\cdot\overline{\Sigma}_{{\rm V}}\frac{1}{2}\epsilon^{\mu\rho\alpha\beta}q_{\rho}f_{\mathrm{f}}^{<}f_{\mathrm{f}}^{>}\Bigr\},\label{eq:MasslessEqui_AWF}
\end{eqnarray}
where $\overline{\Sigma}_{{\rm V}}={\rm Re}\Sigma_{{\rm V}}^{{\rm r}}$.
Similarly, we work out the vector Wigner function up to $\mathcal{O}(\partial^{1})$
from Eq.~(\ref{eq:FWF_Keldysh_eq_1}), 
\begin{eqnarray}
\mathcal{V}^{<,\mu}(q,X) 
 & = & \mathcal{V}_{0}^{<,\mu}+2\pi\epsilon(q_{0})\Bigl[\delta(q^{2})\left(q^{\mu}-\overline{\Sigma}_{{\rm V}}^{\mu}\right)\nonumber \\
 &  & \;-\delta^{\prime}(q^{2})2q^{\mu}q\cdot\overline{\Sigma}_{{\rm V}}\Bigr]f_{{\rm f}}^{<}.
\end{eqnarray}
In the chiral basis, $\mathcal{W}^{\lessgtr,\mu}_{{\rm R/L}}=\mathcal{V}^{\lessgtr,\mu}\pm\mathcal{A}^{\lessgtr,\mu}$, 
$\overline{\Sigma}^{\mu}_{{\rm R/L}}=\overline{\Sigma}^{\mu}_{{\rm V}}$ taking the HTL approximation 
and from Eq.~(\ref{eq:LG_FWF_global_eq}),
\begin{eqnarray}
\mathcal{A}_0^{<,\mu}=2\pi\epsilon(q_{0})\delta(q^{2}-m^{2})f_{\mathrm{f}}^{<}f_{\mathrm{f}}^{>}\frac{1}{4}\epsilon^{\mu\nu\rho\sigma}\Omega_{\nu\rho}q_{\sigma},
\end{eqnarray}
we are able to exactly reproduce the equilibrium chiral Wigner function in Eq.~(\ref{eq:Equi_CWF}) and the modified equilibrium distribution function in Eq.~(\ref{eq:Full_CDist}) without resorting to any knowledge of the modified equilibrium condition in Eq.~(\ref{eq:Geq_cond_2}) or the kinetic equations.

\subsection{Radiative corrections to massless and massive axial vortical effect }

Since we obtain the axial-vector component of Wigner functions in thermal equilibrium with vorticity in the pervious subsection, 
we further evaluate the coupling-dependent
AVE for massive quarks in the thermal QCD background. We also refer to Refs.~\cite{Golkar:2012kb,Hou:2012xg} for related studies on massless fermions, and to Ref.~\cite{Buzzegoli:2021jeh} for investigations of massive fermions under different conditions.

In global equilibrium with pure rotation, from Eq.~(\ref{eq:AV_WF_HTLa}), the axial-charge current is given by 
\begin{eqnarray}
J_{5,\mu}(X) & = & 4\int\frac{{\rm d}^{4}q}{(2\pi)^{4}}{\rm Tr}_{{\rm c}}\mathcal{A}_{\mu}^{<}(q,X)\nonumber \\
 & = & J_{5,\mu}^{0}(X)+\delta J_{5,\mu}(X).
\end{eqnarray}
The zeroth-order result $J_{5,\mu}^{0}(X)$, independent of the interactions, is consistent with the previous finding \citep{Landsteiner:2011cp,Gao:2012ix},
\begin{eqnarray}
J_{\mu}^{5,0} & = & \frac{N_{{\rm c}}T^{2}\omega_{\mu}}{\pi^{2}}I_{0}(\beta m),\;\\
I_{0}(a) & = & \int_{0}^{+\infty}{\rm d}x\frac{(2x^{2}+a^{2})(x^{2}+a^{2})^{-\frac{1}{2}}}{e^{\sqrt{x^{2}+a^{2}}}+1},
\end{eqnarray}
where $I_{0}(0)=\frac{\pi^{2}}{6}$. The $\mathcal{O}(g^{2})$ correction, $\delta J_{5,\mu}(X)$,
reads
\begin{eqnarray}
\delta J_{5}^{\mu}(X) & = & \frac{2N_{c}m_{{\rm f}}^{2}\omega^{\mu}}{\pi^{2}}I_{1}(\beta m),\label{eq:=00005Cdelta-J5-temp1}
\end{eqnarray}
where 
\begin{eqnarray}
 &  & I_{1}(a)\nonumber \\
 & = & \int_{0}^{+\infty}\frac{{\rm d}x}{e^{\sqrt{x^{2}+a^{2}}}+1}\left\{ -\frac{2a^{2}}{x^{3}}\ln\frac{\sqrt{x^{2}+a^{2}}+x}{a}\right.\nonumber \\
 &  & \;+\frac{2(x^{2}+3a^{2})}{3x^{2}\sqrt{x^{2}+a^{2}}}-\frac{e^{\sqrt{x^{2}+a^{2}}}(e^{\sqrt{x^{2}+a^{2}}}-1)}{(e^{\sqrt{x^{2}+a^{2}}}+1)^{2}}\frac{x^{2}}{2\sqrt{x^{2}+a^{2}}}\nonumber \\
 &  & \;\left.+\frac{e^{\sqrt{x^{2}+a^{2}}}}{e^{\sqrt{x^{2}+a^{2}}}+1}\left(1-\frac{2a^{2}+x^{2}}{2x\sqrt{x^{2}+a^{2}}}\ln\frac{\sqrt{x^{2}+a^{2}}+x}{a}\right)\right\} .\nonumber \\
\end{eqnarray}
In the calculation of $\delta J_{5}^{\mu}(X)$, we have inserted the
normal ordering in the definition (\ref{eq:Def_Fermion_WF}) to eliminate
the ultraviolet (UV) divergence from vacuum.

Equation~(\ref{eq:=00005Cdelta-J5-temp1}) is infrared (IR) safe and can be evaluated
numerically for arbitrary nonzero mass since as $x\to0$,
\begin{eqnarray}
-\frac{a^{2}}{x^{3}}\ln\frac{\sqrt{x^{2}+a^{2}}+x}{a}+\frac{x^{2}+3a^{2}}{3x^{2}\sqrt{x^{2}+a^{2}}} & \to & \frac{2x^{2}}{15a^{3}},\\
1-\frac{2a^{2}+x^{2}}{2x\sqrt{x^{2}+a^{2}}}\ln\frac{\sqrt{x^{2}+a^{2}}+x}{a} & \to & 0.
\end{eqnarray}
This property shows that the integral kernel in Eq.~(\ref{eq:=00005Cdelta-J5-temp1})
is non-analytic for $m$ in the small $|\boldsymbol{q}|$ region,
which means that if working in the massless limit, Eq.~(\ref{eq:=00005Cdelta-J5-temp1})
is intrinsically IR divergent. More explicitly, we have
\begin{eqnarray}
 &  & \delta J_{5,m=0}^{\mu}(X)\nonumber \\
 & = & \frac{N_{c}C_{{\rm F}}g^{2}T^{2}\omega^{\mu}}{16\pi^{2}}+\frac{N_{c}m_{{\rm f}}^{2}\omega^{\mu}}{3\pi^{2}}\int_{\Lambda_{\rm IR}}^{+\infty}\frac{{\rm d}|\boldsymbol{q}|}{|\boldsymbol{q}|}f_{{\rm f}}^{<}(|\boldsymbol{q}|),\label{eq:Massless_Keldysh_J5}
\end{eqnarray}
where an IR cutoff $\Lambda_{\rm IR}$ proportional to thermal mass has to be introduced to regularize the result.
However, it is not surprised that Eq.~(\ref{eq:Massless_Keldysh_J5}) does not recover the exact result found in Refs.~\citep{Golkar:2012kb,Hou:2012xg}. Because the HTL approximation
 excludes the energy scale of the phase-space momentum for $|\boldsymbol{q}|\ll T$,
 the integral range cannot cover the whole momentum space. Indeed,
a careful analysis indicates that some $\mathcal{O}(T^{2})$ contributions
are missed in Eq.~(\ref{eq:Massless_Keldysh_J5}). A
diagrammatic comparison between our phase-space formalism and the linear
response theory \citep{Golkar:2012kb,Hou:2012xg} shows that the results without the HTL approximation are indeed equivalent \citep{Fang2:2025}.

\section{Spin alignment of vector mesons from coalescenced quarks \label{sec:Vector-meson-spin}}

Another important application of the Keldysh equation formalism
is to derive the interaction correction to the spin density matrix
of vector mesons. Instead of solving the spin Boltzmann equation for vector-meson
fields \citep{Sheng:2022ffb,Sheng:2022wsy,Kumar:2022ylt,Kumar:2023ghs}
or using the linear response theory \citep{Li:2022vmb,Dong:2023cng}, the Keldysh equation
provides another systematic way to derive the off-equilibrium corrections
of spin density matrix elements. We focus on the boundary of the QGP and hadronic phase, where the self-energies of vector mesons receive corrections from quarks in thermal equilibrium through the effective quark-meson interactions. We also neglect the background strong
(color) fields and concentrate on the contribution of leading hydrodynamic gradients for simplicity. 

The phase-space spin density matrix elements is related to the vector-field Wigner function via the projection by polarization vectors $\epsilon_{\lambda}^{\mu}(q)$ with $\lambda=\pm 1,0$
\citep{Becattini:2020sww,Wagner:2022gza}: 
\begin{eqnarray}
 \rho_{00}(\boldsymbol{q},X)& = & \frac{\epsilon_{0}^{\mu}(q)\epsilon_{0}^{*,\nu}(q)G_{\mu\nu}^{<}(\boldsymbol{q},X)}{\sum_{\lambda}\epsilon_{\lambda}^{\mu}(q)\epsilon_{\lambda}^{*,\nu}(q)G_{\mu\nu}^{<}(\boldsymbol{q},X)}\nonumber \\
 & = & \frac{1}{3}+\delta\rho_{00}(\boldsymbol{q},X). \label{eq:=00005Crho00}
\end{eqnarray}
where 
\begin{eqnarray}
    \delta\rho_{00}(\boldsymbol{q},X) &=& \frac{\left[\epsilon_{0}^{\mu}(q)\epsilon_{0}^{*,\nu}(q)-\frac{1}{3}\sum_{\lambda}\epsilon_{\lambda}^{\mu}(q)\epsilon_{\lambda}^{*,\nu}(q)\right]\delta G_{\mu\nu}^{<}}{\sum_{\lambda}\epsilon_{\lambda}^{\mu}(q)\epsilon_{\lambda}^{*,\nu}(q)G_{\mu\nu,0}^{<}(\boldsymbol{q},X)}. \nonumber \\
\end{eqnarray}
Here, we decompose $G_{\mu\nu}^{<}$ as
\begin{eqnarray}
    G_{\mu\nu}^{<}\approx G_{\mu\nu,0}^{<}+\delta G_{\mu\nu}^{<},
\end{eqnarray}
where $G_{\mu\nu,0}^{<}$ is independent of the interactions and $\delta G_{\mu\nu}^{<}$ is the leading-order interaction-dependent correction.
In the second step of Eq.~(\ref{eq:=00005Crho00}) we have used the expression for the interaction-free equilibrium vector-meson Wigner function up to $\mathcal{O}(\partial^1)$ which is derived from KMS relation \citep{Fang1:2025},
\begin{eqnarray}
 &  & G_{\mu\nu,0}^{<}(q,X)\nonumber \\
 & = & -2\pi\epsilon(q_{0})\delta(q^{2}-m^{2}_V)\left[(\eta_{\mu\nu}-\frac{q_{\mu}q_{\nu}}{m^{2}_V})f_{\mathrm{b}}^{<}(q,X)\right.\nonumber \\
 &  & \;\;\left.+i\left(\frac{q_{[\nu}\Omega_{\mu]\alpha}q^{\alpha}}{m^{2}_V}-\Omega_{\mu\nu}\right)f_{\mathrm{b}}^{<}(q,X)f_{\mathrm{b}}^{>}(q,X)\right],
\end{eqnarray} 
where $m_V$ represents the mass of vector mesons throughout this section.
The polarization four-vector can be explicitly written as  
\begin{eqnarray}
\epsilon_{\lambda}^{\mu}(q)=\Big(\frac{{\bm q}\cdot{\boldsymbol{\epsilon}_{\lambda}}}{m_V},\,\boldsymbol{\epsilon}_{\lambda}+\frac{({\bm q}\cdot\boldsymbol{\epsilon}_{\lambda})\boldsymbol{q}}{m_V(E_{\boldsymbol{q}}^V+m_V)}\Big),
\end{eqnarray}
in terms of $\boldsymbol{\epsilon}_{\lambda}$ as the polarization in the rest frame of the vector meson. The $E_{\boldsymbol{q}}^V$ denotes the energy of vector mesons. The polarization four-vector satisfies the
transverse condition, $\epsilon_{\lambda}\cdot q=0$, polarization sum, 
$\sum_{\lambda}\epsilon_{\lambda}^{\mu}(q)\epsilon_{\lambda}^{*,\nu}(q)=-(\eta^{\mu\nu}-\frac{q^{\mu}q^{\nu}}{q^{2}})$,
and normalization condition, $\epsilon_{\lambda}\cdot\epsilon_{\lambda^{\prime}}=-\delta_{\lambda\lambda^{\prime}}$
\citep{Greiner:1996zu}. In this work, we only discuss the out-of-plane spin
alignment $\boldsymbol{\epsilon}_{0}=(0,1,0)$ along the direction
of global angular momentum \citep{Sheng:2022ffb,Sheng:2022wsy}.  
In addition, the Wigner functions in Eq.~(\ref{eq:=00005Crho00}) are actually on-shell and we should integrate over $q_{0}$ for
$G^<_{\mu\nu}$. In such a case, only the symmetric and real part of $\delta G_{\mu\nu}^{<}$
are relevant since $\epsilon_0^{\mu}$ is real and the polarization contraction $\epsilon_0^{\mu}\epsilon_0^{*\nu}$ in Eq.~(\ref{eq:=00005Crho00}) is symmetric w.r.t. the Lorentz indices. The terms proportional to $\{q_{\mu},q_{\nu},\eta_{\mu\nu}\}$
can be safely neglected since they are orthogonal to $\epsilon_{0}^{\mu}(q)\epsilon_{0}^{*,\nu}(q)-\frac{1}{3}\sum_{\lambda}\epsilon_{\lambda}^{\mu}(q)\epsilon_{\lambda}^{*,\nu}(q)$.
In Eq.~(\ref{eq:=00005Crho00}), we have also neglected the additional second-order gradient corrections from the gradient expansion of the polarization vectors in the Wigner transformation \cite{Kumar:2023ojl}.  Also see Refs.~\cite{Zhang:2024mhs,Yang:2024fkn} for related studies
by Zubarev's approach.

Instead of considering the corrections from the background fields in Sec.\ref{sec:Fermion-spin-polarization},
we focus on the in-medium coalescence and dissociation (leading-order scattering) contribution
to the vector mesons, which means we may approximate the r/a self-energies
up to $\mathcal{O}(\partial^{0})$ as 
\begin{eqnarray}
\Sigma_{\mathrm{r}}^{\mu\nu} & \approx & -\Sigma_{\mathrm{a}}^{\mu\nu}\;\approx\;\frac{i}{2}(\Sigma^{>,\mu\nu}-\Sigma^{<,\mu\nu})\nonumber \\
 & = & i\frac{e^{\beta(q\cdot u-\mu_{{\rm b}})}-1}{2}\Sigma^{<,\mu\nu},
\end{eqnarray}
where in the last step we have used the KMS relation for equilibrium
self-energies \citep{Bellac:2011kqa}. Here $\mu_{{\rm b}}$ denotes the chemical potential of vector mesons. With these preparations, the
non-trivial correction of free vector-meson Wigner function up to
$\mathcal{O}(\partial^{1})$ can be calculated from Eq.~(\ref{eq:BWF_Keldysh_eq}),
\begin{eqnarray}\nonumber
 &&\delta G_{\mu\nu}^{<}
 \\
 & = & -{\rm P.V.}\frac{1}{(q^{2}-m_V^{2})^{2}}\Sigma_{\mu\nu}^{<,0}+2\pi\epsilon(q_{0})\delta^{\prime\prime}(q^{2}-m_V^{2})\nonumber \\
 &  & \;\times\left[-\frac{1}{3}(q^{\lambda}q^{\alpha}\xi_{\lambda\alpha}-q^{\lambda}(\partial_{\lambda}\alpha))\Sigma_{\mu\nu}^{<,0}f_{\mathrm{b}}^{>}-\frac{2}{3}q^{\lambda}\partial_{\lambda}\Sigma_{\mu\nu}^{<,0}\right]\nonumber \\
 &  & +2\pi\epsilon(q_{0})\delta^{\prime}(q^{2}-m_V^{2})\frac{1}{4}f_{\mathrm{b}}^{>}\biggl[-2\Sigma_{\mu\alpha}^{<,0}\Omega_{\;\nu}^{\alpha}+\beta u\cdot\partial\Sigma_{\mu\nu}^{<,0}\nonumber \\
 &  & \;-(\partial_{q}^{\lambda}\Sigma_{\mu\nu}^{<,0})(q^{\alpha}\Omega_{\lambda\alpha}+q^{\alpha}\xi_{\lambda\alpha}-\partial_{\lambda}\alpha)\nonumber \\
 &  & \;+\frac{q^{\alpha}}{m_V^{2}}\Sigma_{\mu\alpha}^{<,0}\left(q^{\lambda}\xi_{\nu\lambda}-\partial_{\nu}\alpha\right)\biggr],\label{eq:NLO_VM_WF}
\end{eqnarray}
where we have used the interaction-free equilibrium form of $G_{\mu\nu}^{<}$ up to the leading
gradient (\ref{eq:App_Less_VMWF_global_eq})
with $f_{{\rm b}}^{<}=f_{{\rm b}}^{>}-1=1/[e^{\beta(u\cdot q-\mu_{{\rm b}})}-1]$
the equilibrium mesonic distribution function. Here $\Sigma_{\mu\nu}^{<,0}$
is the leading-order self-energy with leading-order quark propagators.
Similar properties of the Delta function like Eq.~(\ref{eq:Regularized_delta_property_1})
are used, which can be proven by using its Lorentzian-regularized definition.
Similar to Eq.~(\ref{eq:AV_WF_HTLa}), the corrections (\ref{eq:NLO_VM_WF})
modify the on-shell conditions and possibly deviate the vector mesons
from the quasi-particle picture. 

The first term in Eq.~(\ref{eq:NLO_VM_WF})
denotes the leading order correction to $\rho_{00}$ which can be
derived from standard thermal field theory technique \citep{Bellac:2011kqa}. It also modifies the spectral functions. See  Refs.~\citep{Li:2022vmb,Dong:2023cng} for relevant studies from linear response theory. In this work, we are interested in the the remaining 
terms $\sim\mathcal{O}(\partial^{1})$, which denote the hydrodynamic gradient contributions. From the Keldysh equation for vector-meson fields in Eq.~(\ref{eq:BWF_Keldysh_eq}), there are two possible contributions from Moyal product and $\mathcal{O}(\partial^1)$ self-energy (or Wigner function). It can be shown that the mesonic self-energy $\Sigma_{\mu\nu}$ is symmetric w.r.t. the Lorentz indices in the leading gradient and its $O(\partial^1)$ part is anti-symmetric. 
Inserting the leading-order vector-field Wigner function in Eq.~(\ref{eq:App_Less_VMWF_global_eq}) and using the properties of the polarization vector, we find such a quantum self-energy makes no contribution. Therefore, the spin alignment at $\mathcal{O}(\partial^1)$ only comes from the quantum sector of vector-field Wigner functions and Moyal products.

The self-energies can be explicitly evaluated with the effective quark-meson
interactions as an example. Here, we consider the bare vector vertex for $K^{*}$ \citep{Faessler:2005gd},
\begin{eqnarray}
\mathcal{L}_{qqV}^{\mathrm{vector},K*} & = & g_{\mathrm{V}}\left[\overline{u}\gamma^{\mu}K_{\mu}^{*+}s+\overline{d}\gamma^{\mu}K_{\mu}^{*0}s+{\rm c.c.}\right],
\end{eqnarray}
where $u,d,s$ are the constituent quark fields and c.c. is the charge conjugation of the first two terms. The self-energy
for $K^{*,0}$ reads 
\begin{eqnarray}
 &  & \Sigma_{K*0}^{<,\mu\nu}(q,X)\nonumber \\
 & = & -g_{\mathrm{V}}^{2}\int\frac{\mathrm{d}^{4}p}{(2\pi)^{4}}\theta(q_{0})\mathrm{Tr}\left[\gamma^{\mu}S_{0,d}^{<}(q+p,X)\gamma^{\nu}S_{0,s}^{>}(p,X)\right].\nonumber \\
\label{eq:K*0_SE}
\end{eqnarray}
In RHIC-energies, $e^{\beta m_{s}}\gg1$ with $m_{s}\sim0.42$ GeV being the constitute strange quark mass and
$T\sim0.15$ GeV as the freeze-out temperature, so the distribution of $s$-quark can be approximated
as Boltzmann-type, and analytical expression of Eq.~(\ref{eq:K*0_SE})
can be obtained,
\begin{eqnarray}
\Sigma_{K*0}^{<,\mu\nu}(q,X) & \approx & -\frac{g_{\mathrm{V}}^{2}e^{\beta(-q_{0}+\mu_{K*0})}}{2\pi\beta|\boldsymbol{q}|}\Big[-u^{\mu}u^{\nu}h_{2}(q,X)\nonumber \\
 &  & +(u^{\nu}q^{\mu}+u^{\mu}q^{\nu})h_{1}(q,X)-q^{\mu}q^{\nu}h_{3}(q,X)\Big].\nonumber \\
\label{eq:partial0_VMSE}
\end{eqnarray}
where $\mu_{K^{*0}}$ is the chemical potential of $K^{*0}$ meson
and $h_{1,2,3}$ are defined in Appendix \ref{subsec:Auxiliary-functions-in}.
The gradients of the self-energy can be accordingly evaluated. Then performing the $q_{0}$ integration with $\delta^{(n)}(q^{2}-m^{2})=(2q_{0})^{-1}\partial_{q0}\delta^{(n-1)}(q^{2}-m^{2})$ for Eq.~(\ref{eq:NLO_VM_WF}),
one can finally obtain $\delta\rho_{00}(\boldsymbol{q},X)$. However,
such calculation is rather involved and we only carry out part of the
calculation for illustration and an order-of-magnitude estimate. For example, for the first term after
$\delta^{\prime\prime}(q^{2}-m^{2})$ in Eq.~(\ref{eq:NLO_VM_WF}),
performing integration by parts only on $f_{{\rm b}}^{>}$, namely, 
\begin{eqnarray}
 &  & \int_{0}^{+\infty}\frac{{\rm d}q_{0}}{2\pi}\delta^{\prime\prime}(q^{2}-m_{V}^{2})\times A\times f_{\mathrm{b}}^{>}\nonumber\\
 & \sim & \int_{0}^{+\infty}\frac{{\rm d}q_{0}}{2\pi}\delta(q^{2}-m_{V}^{2})\times A \nonumber\\
 &  & \times f_{\mathrm{b}}^{>}\left[\frac{3\beta}{4q_{0}^{3}}(f_{\mathrm{b}}^{<}+\frac{T}{q_{0}})+\frac{\beta^{2}}{4q_{0}^{2}}\left(1+2f_{\mathrm{b}}^{<}\right)f_{\mathrm{b}}^{<}\right]
\end{eqnarray}
with $A=-\frac{1}{3}(q^{\lambda}q^{\alpha}\xi_{\lambda\alpha}-q^{\lambda}(\partial_{\lambda}\alpha))\Sigma_{\mu\nu}^{<,0}$,  its modification to the density operator reads
\begin{eqnarray}
 &  & \delta\rho_{00}^{(2,1),y}(\boldsymbol{q},X)\nonumber \\
 & = & \sum_{i=x,y,z}\left[I_{\nabla\alpha}^{(2,1),i}\beta\boldsymbol{\nabla}^{i}\alpha+I_{\nabla\beta}^{(2,1),i}(\boldsymbol{\nabla}^{i}\beta-(u\cdot\partial)\boldsymbol{u}^{i})\right]\nonumber \\
 &  & +I_{D\alpha}^{(2,1)}\beta(u\cdot\partial)\alpha+I_{D\beta}^{(2,1)}(u\cdot\partial)\beta+I_{\theta}\beta(\partial\cdot u)\nonumber \\
 &  & +\sum_{i,j=x,y,z}I_{\sigma}^{(2,1),ij}\beta\sigma^{ij},
\end{eqnarray}
where $I^{(2,1),i}$ are some complicated momentum-dependent coefficients with subscript 2,1 representing the contribution from the first part after the integration by parts for the term attached to the second-order derivative of the Delta function in Eq.~(\ref{eq:NLO_VM_WF}) and $\sigma^{\mu\nu}=\Delta^{\mu(\alpha}\Delta^{\beta)\nu}\partial_{\alpha}u_{\beta}-\frac{1}{3}\Delta^{\mu\nu}(\partial\cdot{u})$
 is the shear-stress tensor. 

For an order-of-magnitude estimate, in $\sqrt{s_{\mathrm{NN}}}=19.6$ GeV, we
choose the parameters, $\mu_{d}\simeq\mu_{K*0}=0.07$ GeV \citep{STAR:2017sal},
$T=0.155$ GeV, and $m_{s}=0.419$ GeV, $m_{d}=0.220$ GeV \citep{Godfrey:1985xj},
$m_{K*0}=0.895$ GeV, and $g_{\mathrm{V}}=1$. By considering the
transverse momentum region $1.2<q_{\mathrm{T}}<5.4$ GeV, the rapidity
region $-1<Y<+1$ and the full range of azimuthal angle $0<\varphi<2\pi$
\citep{STAR:2022fan}, we find only the even moment of $\boldsymbol{q}^{i}$
survives,
\begin{eqnarray}
\langle I_{D\alpha}^{(2,1)}\rangle & = & \frac{\int\mathrm{d}^{3}\boldsymbol{q}I_{D\alpha}(\boldsymbol{q})f^{<}(q,X)}{\int\mathrm{d}^{3}\boldsymbol{q}f^{<}(q,X)}=-6.34\times10^{-4},
\end{eqnarray}
and 
\begin{eqnarray}
\langle I_{D\beta}^{(2,1)}\rangle=6.78\times10^{-3} & ,\; & \langle I_{\theta}^{(2,1)}\rangle=-4.79\times10^{-3},\nonumber \\
\langle I_{\sigma}^{(2,1),xx}\rangle=-1.90\times10^{-3} & ,\; & \langle I_{\sigma}^{(2,1),yy}\rangle=6.48\times10^{-3},\nonumber \\
\langle I_{\sigma}^{(2,1),zz}\rangle=2.02\times10^{-4} & .
\end{eqnarray}
We expect the relevant coefficients as the counterpart of $I^{(2,1),i}$ for the remaining terms in Eq.~(\ref{eq:NLO_VM_WF}) will be of the same order of magnitude. Similar calculations for $\phi$ mesons show the coefficients are
in general larger than that of the $K^{*0}$ mesons in magnitude. On the other
hand, the gradient of hydrodynamic variables are of $\mathcal{O}(10^{-2})$
in general, so these contributions are further suppressed by two orders of magnitude
and do not significantly modify the global spin alignment of mesons,
which is consistent with results from linear response theory \citep{Dong:2023cng}. 

\section{Summary and outlook \label{sec:Summary} }

We have investigated the radiative corrections to the spin-polarization spectrum and the AVE for massive fermions in a hot QCD background. 
To resolve the
difficulty in determining the interaction corrections to Wigner functions
up to the leading-order coupling, we formulate the Keldysh-equation framework, which is shown in Eq.~(\ref{eq:Fermion_Keldysh_Eq_1}) for fermions and Eq.~(\ref{eq:BWF_Keldysh_eq}) for bosons. These Keldysh equations are the integral-equation parallel of the Kadanoff-Baym differential-integral equations and the generalization of Dyson-Schwinger equation to the phase space in off-equilibrium system and captures the non-perturbative features. In principle, it can be applicable to a strongly coupled system and large hydrodynamic gradient. 

Up to the leading order in gradients, using the iteration method, we have recovered the results based on QKT and determined the interaction correction to a global
equilibrium axial distribution function in the lowest order of coupling constant. We present how  to add the background EMF to our theoretical framework in a systematic way. We further compute the complete lowest order global equilibrium corrections to the spin polarization pseudo-vector from a QCD medium based on the HTL approximation in Eqs.~(\ref{eq:C-F_formulae_therm}, \ref{eq:C-F_formulae_acc}, \ref{eq:C-F_formulae_vor}), which also include the contributions from the previously omitted dynamical sector in QKT. Such contributions can also impact on the global polarization of hyperons. Its local equilibrium generalization is also attainable from Eq.~(\ref{eq:Fermion_Keldysh_Eq_1}) if local equilibrium $\mathcal{O}(\partial^1)$ fermionic and gluonic Wigner functions are given.
 
We also compute the radiative corrections to both massless and massive AVE. Such corrections of the AVE for massive fermions  in our study cannot reproduce the massless case in a smooth manner due to the illness of the infrared structure for HTL self-energies.  Nonetheless, a diagrammatic comparison shows that such one-loop self-energy is indeed the origin
of radiative corrections found in Refs.~\citep{Golkar:2012kb,Hou:2012xg}, which will be presented elsewhere \citep{Fang2:2025}.

At last, we evaluate $\delta\rho_{00}$
 under the effective quark-meson interactions based on the Keldysh-equation formalism.  Our result is indeed consistent with linear response theory \citep{Dong:2023cng}. As a complementary study, we will present the explicit calculation for self-energy corrected
fermionic Wigner function from the linear response theory in an upcoming
paper \citep{Fang2:2025}. 

Before ending this work, we would like to remark on the limitation of the framework based on the Keldysh-equation. We emphasize that Keldysh-equation itself does not contain the complete information of the physical system. As shown in this work,  it is necessary to incorporate the non-interacting Wigner functions derived from its equations of motion as inputs. A similar strategy is widely used in early works on quantum kinetic theory (see, e.g., Refs.~\cite{Gao:2012ix, Chen:2012ca}), in which the non-interacting Wigner functions at the leading order of $\hbar^0$ are assumed based on classical kinetic theory.

\begin{acknowledgments}
We would like to thank Fei Gao and Shu Lin for helpful discussions.
This work is supported in part by the National Key Research and Development
Program of China under Contract No. 2022YFA1605500, by the Chinese
Academy of Sciences (CAS) under Grant No. YSBR-088 and by National Natural Science Foundation of China (NSFC) under Grants No. 12135011. S. Fang is supported in part
by the China Scholarship Council (CSC) when finishing this manuscript. D.-L. Y. is supported by National Science and Technology Council (Taiwan) under Grants No. NSTC 113-2628-M-001-009-MY4 and by Academia Sinica under Project No. AS-CDA-114-M01. 
\end{acknowledgments}

\appendix

\section{Non-interacting equilibrium Wigner functions \label{subsec:Non-interacting-equilibrium-Wign}}

We present the Wigner functions for spin-half and spin-1 particles
in global equilibrium up to $\mathcal{O}(\partial^{1})$, which can
be diagrammatically derived by expanding the global equilibrium density operator up
to the linear order \citep{Becattini:2013fla,Buzzegoli:2017cqy,Palermo:2021hlf,Fang2:2025}
or from a generalized KMS relation \citep{Fang1:2025}. For
fermions,
\begin{eqnarray}
 &  & S_{0}^{\lessgtr}(q,X)\nonumber \\
 & = & 2\pi\epsilon(q_{0})\delta(q^{2}-m^{2})\left[(\gamma^{\mu}q_{\mu}+m)f_{\mathrm{f}}^{\lessgtr}(q,X)\right.\nonumber \\
 &  & \left.\pm f_{\mathrm{f}}^{<}(q,X)f_{\mathrm{f}}^{>}(q,X)\frac{1}{4}\Omega_{\nu\rho}\left(\epsilon^{\mu\nu\rho\sigma}\gamma^{5}\gamma_{\mu}q_{\sigma}-m\sigma^{\nu\rho}\right)\right],\nonumber \\
\label{eq:LG_FWF_global_eq}
\end{eqnarray}
where $f_{\mathrm{f}}^{<}(q,X)= 1/[e^{\beta(u\cdot q-\mu)}+1]$. In this work we work in a near-equilibrium limit, i.e. the distribution
functions do not vanish the Boltzmann equations, so the configuration
of $u^{\mu}$ can be general. It is noticed that for the fermionic field,
the terms in lesser/greater Wigner functions of $\mathcal{O}(\partial^{n})$,
$n\geq1$ are only different up to a minus sign. The r/a quantities
are related with each other via
\begin{eqnarray}
S^{\mathrm{r/a}}(q,X) & = & \int\frac{\mathrm{d}q_{0}^{\prime}}{2\pi}\frac{\rho(q_{0}^{\prime},\boldsymbol{q};X)}{q_{0}^{\prime}-q_{0}\mp i\eta},\\
\Sigma^{\mathrm{r/a}}(q,X) & = & \int\frac{\mathrm{d}q_{0}^{\prime}}{2\pi}\frac{\Gamma(q_{0}^{\prime},\boldsymbol{q};X)}{q_{0}^{\prime}-q_{0}\mp i\eta}.
\end{eqnarray}
The spectral function $\rho$ and width function $\Gamma$ are defined
as
\begin{eqnarray}
\rho=S^{>}+S^{<} & ,\; & \Gamma=\Sigma^{>}+\Sigma^{<}.
\end{eqnarray}
where the sign before the lessor quantities are opposite to the bosonic
case due to statistical properties. For the free equilibrium Wigner
function \citep{Bellac:2011kqa},
\begin{eqnarray}
\rho(q,X) & = & 2\pi\epsilon(q_{0})\delta(q^{2}-m^{2})(\gamma^{\mu}q_{\mu}+m),
\end{eqnarray}
and the r/a Wigner functions are easily calculated,
\begin{eqnarray}
S_{0}^{\mathrm{r/a}}(q,X) & = & -\frac{\gamma^{\mu}q_{\mu}+m}{q^{2}-m^{2}\pm iq_{0}\eta}.\label{eq:r/a_FWF_global_eq}
\end{eqnarray}
which is spacetime-independent as expected. Especially, the scalar components read $\mathcal{F}_{0}^{\mathrm{r/a}}(q,X)  =  -\frac{m}{q^{2}-m^{2}\pm iq_{0}\eta}$.
All other Wigner functions, like (anti-)time-order,
can be expressed in terms of the r/a and Wightman Wigner functions.

Similarly for the massive vector field,
\begin{eqnarray}
 &  & G_{0,\mu\nu}^{<}(q,X)\nonumber \\
 & = & -2\pi\epsilon(q_{0})\delta(q^{2}-m^{2})\left[(\eta_{\mu\nu}-\frac{q_{\mu}q_{\nu}}{m^{2}})f_{\mathrm{b}}^{<}(q,X)\right.\nonumber \\
 &  & \;\;\left.+i\left(\frac{q_{[\nu}\Omega_{\mu]\alpha}q^{\alpha}}{m^{2}}-\Omega_{\mu\nu}\right)f_{\mathrm{b}}^{<}(q,X)f_{\mathrm{b}}^{>}(q,X)\right],\label{eq:App_Less_VMWF_global_eq}
\end{eqnarray}
where the bosonic distribution function is defined as 
\begin{eqnarray}
f_{\mathrm{b}}^{<}(q,X) & = & \frac{1}{e^{\beta u\cdot q-\mu}-1}.
\end{eqnarray}
For the bosonic field, opposite to the fermionic case, the terms in
lesser/greater Wigner functions of $\mathcal{O}(\partial^{n})$, $n\geq1$
are same. And the r/a Wigner functions read
\begin{eqnarray}
G_{0,\mu\nu}^{\mathrm{r/a}}(q,X) & = & \frac{\eta_{\mu\nu}-\frac{q_{\mu}q_{\nu}}{m^{2}}}{q^{2}-m^{2}\pm i\eta q_{0}},\label{eq:r/a_VMWF_global_eq}
\end{eqnarray}
where we have dropped the normal-dependent term which does not contribute
to the calculation of Feynman diagrams \citep{Greiner:1996zu}. 

For the gauge field, the Wigner functions depend on the gauge choice
\citep{Fang1:2025},
\begin{eqnarray}
G_{\mu\nu}^{<}(q,X) & = & 2\pi\epsilon(q_{0})\delta(q^{2})\left[P_{\mu\nu}^{(\xi)}(q)f_{\mathrm{b}}^{<}(q,X)\right.\nonumber \\
 &  & \;\;\left.-2iP_{[\nu}^{(\xi),\alpha}\Omega_{\mu]\alpha}f_{\mathrm{b}}^{<}(q,X)f_{\mathrm{b}}^{>}(q,X)\right],
\end{eqnarray}
where $P_{\mu\nu}^{(\xi)}(q)$ is the polarization summation replying
on the gauge choice \citep{Greiner:1996zu},
\begin{eqnarray}
\sum_{\lambda=1}^{2}\epsilon^{\mu}(\boldsymbol{q},\lambda)\epsilon^{*,\nu}(\boldsymbol{q},\lambda) & = & P_{(\xi)}^{\mu\nu}(q).
\end{eqnarray}
for example, in the Feynman gauge $P_{(\xi)}^{\mu\nu}(q)=-\eta^{\mu\nu}$.
The r/a Wigner functions read
\begin{eqnarray}
G_{\mu\nu,0}^{\mathrm{r/a}}(q,X) & = & \frac{\eta_{\mu\nu}}{q^{2}\pm i\eta q_{0}}.
\end{eqnarray}
\section{The auxiliary functions}

\subsection{Auxiliary functions in spin Cooper-Frye formulae \label{sec:Auxiliary-functions-in}}

We list some auxiliary functions appearing in Eqs.(\ref{eq:C-F_formulae_therm}-\ref{eq:C-F_formulae_vor})
\begin{eqnarray}
 &  & G_{{\rm T}}(\boldsymbol{q})\nonumber \\
 & = & \frac{2T}{|\boldsymbol{q}|}\left(Q_{0}(\frac{E_{\boldsymbol{q}}}{|\boldsymbol{q}|})+\frac{E_{\boldsymbol{q}}}{|\boldsymbol{q}|}Q_{1}(\frac{E_{\boldsymbol{q}}}{|\boldsymbol{q}|})\right)f_{\mathrm{f}}^{<}(E_{\boldsymbol{q}})\nonumber \\
 &  & +\left[\frac{1}{2}+\frac{E_{\boldsymbol{q}}^{2}}{|\boldsymbol{q}|^{2}}Q_{1}(\frac{E_{\boldsymbol{q}}}{|\boldsymbol{q}|})+2\frac{E_{\boldsymbol{q}}}{|\boldsymbol{q}|}R_{1}(\frac{E_{\boldsymbol{q}}}{|\boldsymbol{q}|})\right]f_{\mathrm{f}}^{<}(E_{\boldsymbol{q}})f_{\mathrm{f}}^{>}(E_{\boldsymbol{q}})\nonumber \\
 &  & +\left(\frac{\beta E_{\boldsymbol{q}}}{2}+\frac{\beta E_{\boldsymbol{q}}^{2}}{4|\boldsymbol{q}|}R_{1}(\frac{E_{\boldsymbol{q}}}{|\boldsymbol{q}|})\right)f_{\mathrm{f}}^{<}(E_{\boldsymbol{q}})f_{\mathrm{f}}^{>}(E_{\boldsymbol{q}})\left(1-2f_{\mathrm{f}}^{<}(E_{\boldsymbol{q}})\right),\nonumber \\
\end{eqnarray}
and 
\begin{eqnarray}
 &  & G_{{\rm a}}(\boldsymbol{q})\nonumber \\
 & = & \left(\frac{6E_{\boldsymbol{q}}}{|\boldsymbol{q}|}Q_{0}(\frac{E_{\boldsymbol{q}}}{|\boldsymbol{q}|})+\frac{6E_{\boldsymbol{q}}^{2}}{|\boldsymbol{q}|^{2}}Q_{1}(\frac{E_{\boldsymbol{q}}}{|\boldsymbol{q}|})\right)f_{\mathrm{f}}^{<}(E_{\boldsymbol{q}})\nonumber \\
 &  & +\left(\frac{6\beta E_{\boldsymbol{q}}^{2}}{|\boldsymbol{q}|}R_{1}(\frac{E_{\boldsymbol{q}}}{|\boldsymbol{q}|})+\frac{2\beta E_{\boldsymbol{q}}^{3}}{|\boldsymbol{q}|^{2}}Q_{1}(\frac{E_{\boldsymbol{q}}}{|\boldsymbol{q}|})\right)f_{\mathrm{f}}^{<}(E_{\boldsymbol{q}})f_{\mathrm{f}}^{>}(E_{\boldsymbol{q}})\nonumber \\
 &  & +\frac{\beta^{2}E_{\boldsymbol{q}}^{3}}{|\boldsymbol{q}|}R_{1}(\frac{E_{\boldsymbol{q}}}{|\boldsymbol{q}|})f_{\mathrm{f}}^{<}(E_{\boldsymbol{q}})f_{\mathrm{f}}^{>}(E_{\boldsymbol{q}})\left(1-2f_{\mathrm{f}}^{<}(E_{\boldsymbol{q}})\right),
\end{eqnarray}
and 
\begin{eqnarray}
 &  & G_{{\rm vor}1}(\boldsymbol{q})\nonumber \\
 & = & \left[\frac{4E_{\boldsymbol{q}}}{|\boldsymbol{q}|}Q_{0}(\frac{E_{\boldsymbol{q}}}{|\boldsymbol{q}|})+\frac{2E_{\boldsymbol{q}}^{2}}{|\boldsymbol{q}|^{2}}Q_{1}(\frac{E_{\boldsymbol{q}}}{|\boldsymbol{q}|})+2\right]f_{\mathrm{V}}^{<}(E_{\boldsymbol{q}})\nonumber \\
 &  & +\beta E_{\boldsymbol{q}}\left(1+\frac{E_{\boldsymbol{q}}^{2}+2m^{2}}{|\boldsymbol{q}|^{2}}Q_{1}(\frac{E_{\boldsymbol{q}}}{|\boldsymbol{q}|})+4\frac{E_{\boldsymbol{q}}}{|\boldsymbol{q}|}R_{1}(\frac{E_{\boldsymbol{q}}}{|\boldsymbol{q}|})\right)\nonumber \\
 &  & \;\times f_{\mathrm{f}}^{<}(E_{\boldsymbol{q}})f_{\mathrm{f}}^{>}(E_{\boldsymbol{q}})\nonumber \\
 &  & +\frac{\beta^{2}E_{\boldsymbol{q}}^{3}}{2|\boldsymbol{q}|}R_{1}(\frac{E_{\boldsymbol{q}}}{|\boldsymbol{q}|})f_{\mathrm{f}}^{<}(E_{\boldsymbol{q}})f_{\mathrm{f}}^{>}(E_{\boldsymbol{q}})\left(1-2f_{\mathrm{f}}^{<}(E_{\boldsymbol{q}})\right),
\end{eqnarray}
and 
\begin{eqnarray}
 &  & G_{{\rm vor}2}(\boldsymbol{q})\nonumber \\
 & = & 6\left(\frac{E_{\boldsymbol{q}}}{|\boldsymbol{q}|}Q_{0}(\frac{E_{\boldsymbol{q}}}{|\boldsymbol{q}|})+\frac{E_{\boldsymbol{q}}^{2}}{|\boldsymbol{q}|^{2}}Q_{1}(\frac{E_{\boldsymbol{q}}}{|\boldsymbol{q}|})\right)f_{\mathrm{f}}^{<}(E_{\boldsymbol{q}})\nonumber \\
 &  & +\frac{6\beta E_{\boldsymbol{q}}^{2}}{|\boldsymbol{q}|}Q_{0}(\frac{E_{\boldsymbol{q}}}{|\boldsymbol{q}|})f_{\mathrm{f}}^{<}(E_{\boldsymbol{q}})f_{\mathrm{f}}^{>}(E_{\boldsymbol{q}})\nonumber \\
 &  & +\frac{3\beta^{2}E_{\boldsymbol{q}}^{3}}{|\boldsymbol{q}|}R_{1}(\frac{E_{\boldsymbol{q}}}{|\boldsymbol{q}|})f_{\mathrm{f}}^{<}(E_{\boldsymbol{q}})f_{\mathrm{f}}^{>}(E_{\boldsymbol{q}})\left(1-2f_{\mathrm{f}}^{<}(E_{\boldsymbol{q}})\right),\nonumber \\
\end{eqnarray}
and 
\begin{eqnarray}
 &  & G_{{\rm vor}3}(\boldsymbol{q})\nonumber \\
 & = & \frac{2E_{\boldsymbol{q}}^{2}}{|\boldsymbol{q}|^{2}}\left(\frac{3E_{\boldsymbol{q}}^{2}}{|\boldsymbol{q}|^{2}}Q_{1}(\frac{E_{\boldsymbol{q}}}{|\boldsymbol{q}|})-1\right)f_{\mathrm{f}}^{<}(E_{\boldsymbol{q}})\nonumber \\
 &  & +\frac{\beta E_{\boldsymbol{q}}^{3}}{|\boldsymbol{q}|^{2}}\left(\frac{E_{\boldsymbol{q}}}{|\boldsymbol{q}|}R_{1}(\frac{E_{\boldsymbol{q}}}{|\boldsymbol{q}|})-\frac{2m^{2}}{|\boldsymbol{q}|^{2}}Q_{1}(\frac{E_{\boldsymbol{q}}}{|\boldsymbol{q}|})\right)f_{\mathrm{f}}^{<}(E_{\boldsymbol{q}})f_{\mathrm{f}}^{>}(E_{\boldsymbol{q}}),\nonumber \\
\end{eqnarray}
where we have introduced 
\begin{eqnarray}
R_{1}(x) & = & Q_{0}(x)-xQ_{1}(x).
\end{eqnarray}

\subsection{Auxiliary functions in the vector-meson self-energies \label{subsec:Auxiliary-functions-in}}

We list the auxiliary functions appeared in vector-meson self-energy
(\ref{eq:partial0_VMSE}):

\begin{eqnarray}
h_{1} & = & \left(\frac{q_{0}(q^{2}+3m_{s}^{2}-m_{d}^{2})}{2|\boldsymbol{q}|^{2}}+3q_{0}\frac{(q^{2}+m_{s}^{2}-m_{d}^{2})^{2}}{4|\boldsymbol{q}|^{4}}\right)H_{1}(q_{0})\nonumber \\
 &  & +\left(\frac{m_{s}^{2}-m_{d}^{2}}{|\boldsymbol{q}|^{2}}-\frac{3q_{0}^{2}(q^{2}+m_{s}^{2}-m_{d}^{2})}{|\boldsymbol{q}|^{4}}\right)\frac{H_{2}(q_{0})}{\beta}\\
 &  & +3q_{0}\frac{q^{2}}{|\boldsymbol{q}|^{4}}\frac{H_{3}(q_{0})}{\beta^{2}},\nonumber \\
h_{2} & = & \left((3q_{0}^{2}-|\boldsymbol{q}|^{2})\frac{(q^{2}+m_{s}^{2}-m_{d}^{2})^{2}}{4|\boldsymbol{q}|^{4}}+m_{s}^{2}\frac{q^{2}}{|\boldsymbol{q}|^{2}}\right)H_{1}(q_{0})\nonumber \\
 &  & -\frac{3q^{2}}{|\boldsymbol{q}|^{4}}q_{0}(q^{2}+m_{s}^{2}-m_{d}^{2})\frac{H_{2}(q_{0})}{\beta}+\frac{3q^{4}}{|\boldsymbol{q}|^{4}}\frac{H_{3}(q_{0})}{\beta^{2}},\\
h_{3} & = & \left(\frac{3(q^{2}+m_{s}^{2}-m_{d}^{2})^{2}}{4|\boldsymbol{q}|^{4}}+\frac{q^{2}+2m_{s}^{2}-m_{d}^{2}}{|\boldsymbol{q}|^{2}}\right)H_{1}(q_{0})\nonumber \\
 &  & +\left(-\frac{3q_{0}(q^{2}+m_{s}^{2}-m_{d}^{2})}{|\boldsymbol{q}|^{4}}-\frac{2q_{0}}{|\boldsymbol{q}|^{2}}\right)\frac{H_{2}(q_{0})}{\beta}\nonumber \\
 &  & +\frac{3q_{0}^{2}-|\boldsymbol{q}|^{2}}{|\boldsymbol{q}|^{4}}\frac{H_{3}(q_{0})}{\beta^{2}},
\end{eqnarray}
where the distribution function integral gives

\begin{eqnarray*}
H_{1}(q_{0}) & = & -\beta m_{s}+\beta m_{s}\ln(e^{\beta(q_{0}-\mu_{d}-m_{s})}+1),\\
H_{2}(q_{0}) & = & \beta m_{s}\ln(e^{\beta(q_{0}-\mu_{d}-m_{s})}+1)-\mathrm{Li}_{2}(-e^{\beta(q_{0}-\mu_{d}-m_{s})}),\\
H_{3}(q_{0}) & = & (\beta m_{s})^{2}\ln(e^{\beta(q_{0}-\mu_{d}-m_{s})}+1)\\
 &  & -2\beta m_{s}\mathrm{Li}_{2}(-e^{\beta(q_{0}-\mu_{d}-m_{s})})-2\mathrm{Li}_{3}(-e^{\beta(q_{0}-\mu_{d}-m_{s})}),
\end{eqnarray*}
with $\mathrm{Li}_{n}(z)$ being the poly-logarithm functions and
$\mu_{d}$ the chemical potential of $d$-quark.
\bibliographystyle{h-physrev}
\bibliography{qkt-ref}
 
\end{document}